\documentclass{JHEP3}
\usepackage{epsf}
\usepackage[utf8]{inputenc}
\usepackage{amstext}
\usepackage{graphicx}
\usepackage{amssymb}
\usepackage{esint}
\newcommand{\tOD}{\hat{\Omega}_{D}}
\newcommand{\tOM}{\hat{\Omega}_m}
\newcommand{\tOMo}{\hat{\Omega}_m^0}
\newcommand{\rs}{\rho_{m s}}
\newcommand{\rD}{\rho_D}
\newcommand{\OM}{\Omega_m}
\newcommand{\OL}{\Omega_{\Lambda}}
\newcommand{\OR}{\Omega_R}
\newcommand{\ORo}{\Omega_R^0}

\newcommand{\OKo}{\Omega_K^0}
\newcommand{\rc}{\rho_c}

\newcommand{\rmm}{\rho_{m}}
\newcommand{\rmo}{\rho_{m}^0}
\newcommand{\omm}{\omega_m}
\newcommand{\OLo}{\Omega_{\Lambda}^0}
\newcommand{\OMo}{\Omega_{m}^0}
\newcommand{\rco}{\rho_{c}^0}
\newcommand{\mincir}{\raise
-3.truept\hbox{\rlap{\hbox{$\sim$}}\raise4.truept\hbox{$<$}\ }}
\newcommand{\magcir}{\raise
-3.truept\hbox{\rlap{\hbox{$\sim$}}\raise4.truept\hbox{$>$}\ }}
\newcommand{\be}{\begin{equation}}
\newcommand{\ee}{\end{equation}}
\newcommand{\rL}{\rho_{\Lambda}}

\newcommand{\rLe}{\rho_{\Lambda{\rm eff}}}

\newcommand{\rLo}{\rho_{\Lambda}^0}
\newcommand{\CC}{\Lambda}
\newcommand{\rM}{\rho_m}
\newcommand{\pM}{p_m}

\newcommand{\weff}{\omega_{\rm eff}}

\newcommand{\newtext}[1]{\text{#1}}

\title{Hubble expansion and structure formation in the ``running FLRW model'' of the cosmic evolution}

\author{%
Javier Grande$^a$, Joan Solà$^a$\\
$^a$High Energy Physics Group, Dept.\ ECM, and Institut
de Ciències del Cosmos,\\
Univ.\ de Barcelona, Av.\ Diagonal 647, E-08028 Barcelona,
Catalonia, Spain}

\author{Spyros Basilakos$^{b,a}$, Manolis Plionis$^c$\\
$^{b}$ Academy of Athens, Research Center for Astronomy and Applied
Mathematics, Soranou Efesiou 4, 11527, Athens, Greece

$^{c}$ Institute of Astronomy \& Astrophysics, National Observatory
of Athens,\\ Thessio 11810, Athens, Greece, and Instituto Nacional
de Astrof\'isica, \'Optica y Electr\'onica, 72000 Puebla, Mexico

E-Mails:
\email{jgrande@ecm.ub.es},\email{sola@ecm.ub.es},\email{svasil@academyofathens.gr},\email{mplionis@astro.noa.gr}}

\abstract{%
A new class of Friedmann-Lema\^\i tre-Robertson-Walker (FLRW)
cosmological models with time-evolving fundamental parameters should
emerge naturally from a description of the expansion of the universe
based on the first principles of quantum field theory and string
theory. Within this general paradigm, one expects that both the
gravitational Newton's coupling $G$ and the cosmological term $\CC$
should not be strictly constant but appear rather as smooth
functions of the Hubble rate $H(t)$.
This scenario (``running FLRW model'') predicts, in a natural way,
the existence of dynamical dark energy without invoking the
participation of extraneous scalar fields.
In this paper, we perform a detailed study of some of these models
in the light of the latest cosmological data, which serves to
illustrate the phenomenological viability of the new dark energy
paradigm as a serious alternative to the traditional scalar field
approaches. By performing a joint likelihood analysis of the recent
supernovae type Ia data (SNIa), the Cosmic Microwave Background
(CMB) shift parameter, and the Baryonic Acoustic Oscillations (BAOs)
traced by the Sloan Digital Sky Survey (SDSS), we put tight
constraints on the main cosmological parameters. Furthermore, we
derive the theoretically predicted dark-matter halo mass function
and the corresponding redshift distribution of cluster-size halos
for the ``running'' models studied. Despite the fact that these
models closely reproduce the standard $\CC$CDM Hubble expansion,
their normalization of the perturbation's power-spectrum varies,
imposing, in  many cases, a significantly different cluster-size
halo redshift distribution. This fact indicates that it should be
relatively easy to distinguish between the ``running'' models and
the $\CC$CDM using realistic future X-ray and Sunyaev-Zeldovich
cluster surveys.
}
 \preprint{}

\begin{document}

\section{Introduction}

The high-quality observations performed during the last two decades
have enabled cosmologists to gain substantial confidence that modern
cosmology is capable of quantitatively reproducing the details of
the many observed cosmic phenomena\,\cite{Teg04,Hic09,komatsu08},
including the accelerated expansion of the universe at the present
epoch \cite{SNIa}. The field of cosmology, therefore, is no longer a
pure realm of philosophical speculation, but a rigorous branch of
experimental physics. In view of this fact, the status of modern
cosmology can be declared as very healthy and highly satisfactory
from the point of view of the empirical sciences. Nevertheless, if
we pause for a moment and make a deeper reflection, we soon come to
realize that the ultimate understanding of these magnificent
observations is still highly blurred, if not completely opaque, to
the light of the first principles of fundamental theoretical
physics, e.g. from the viewpoint of Quantum Field Theory (QFT) and
string theory.

Indeed, we realize that what we have been able to do throughout the
last decade is to furnish a rather precise numerical fit to the
parameters of the Friedmann-Lema\^\i tre-Robertson-Walker (FLRW)
model in the light of the most accurate observations, but at the
moment these observational successes are not accompanied by a deeper
theoretical insight into the meaning of the preciously collected
data on dark matter (DM) and dark energy (DE). Needless to say, the
resulting high quality of this combined fit (from independent data
sources) speaks very well for the FLRW model as a likely correct
starting point to understand the structure of our cosmos. However,
it does not provide a single clue to the meaning of -- or on the
physical substratum lying underneath -- the fitted energy density
parameters at the present time $\OMo$ and $\OLo$ (except for the
baryonic part $\Omega_b^0\simeq 0.04$, of course, which we know from
primordial nucleosynthesis), nor does it explain why the geometric
parameter $\OKo$ is comparatively much smaller. In practice,
observations tell us that $\OMo\simeq 0.28$ and $\OLo\simeq 0.72$
(with $\OKo\lesssim 0.01$. It follows that most of the matter
content of the universe is non-baryonic DM, and that the value of
the cosmological constant density $\rL^0=\Lambda/(8\pi G)$, or, in
general, the current dark energy (DE) density, is
\begin{equation}\label{CCvalue}
\rLo=\OL^0\,\rc^0\simeq 6\,h^2\times 10^{-47}\,\rm{GeV}^{4}\simeq 3\times
10^{-47}\,\rm{GeV}^{4}\,,
\end{equation}
where $\rc^0 \simeq \left( 3.0\,\sqrt{h}\times
10^{-12}\,\rm{GeV}\right) ^{4}\sim 4\times 10^{-47}$ GeV$^4$ is the
present value of the critical density and $H_0 = 100\,h$ km/s/Mpc,
with $h \simeq 0.704$, the current Hubble's rate.

Despite the ``scandal'' of our having to admit full ignorance about
the identity {of most ($\sim 90\%$)} of the matter budget of the
universe, this trouble pales in comparison to the colossal enigma
hidden behind the value (\ref{CCvalue}). Understanding this value is
perhaps the biggest scientific mystery of all times: ``the
cosmological constant problem''\,\cite{weinberg89,CCproblem}, which
manifests itself as the double conundrum  of the tiny observed value
of the cosmological constant (CC) in Einstein's equations -- the
``old CC problem''\,\cite{Zeldovich67} --, and also the puzzling
fact that this value is so close to the current matter density (the
``cosmic coincidence problem''\,\cite{Steinhardt}), including the
mystery of a possible phantom behavior near our time. Solving these
problems surely requires introducing new fundamental ideas of
profound theoretical meaning and scope. Some promising new avenues
have recently been proposed\,\cite{BSS10a,BSS10b,
BSS09,SS12,LXCDM}\footnote{For a recent review of some of these
ideas, see \cite{JSP11a} and references therein.} -- see also the
recent alternative
attempts\,\cite{Woodard10,Maggiore10,CCrecent1,CCrecent2}.

For a long time model builders have pursued an explanation of these
cosmological puzzles in terms of dynamical scalar
fields\,\cite{OldScalar}, later adapted into the idea of
quintessence\,\cite{Quintessence}. Quintessence was proposed without
attempting to explain the smallness of the CC, but only to cope with
some aspects of the cosmic coincidence problem\,\cite{CCproblem}.
These scalar field models ignore \textit{ab initio} the huge vacuum
energy from the Standard Model (SM) of strong and electroweak
interactions, which is $55$ orders of magnitude larger than the
measured value (\ref{CCvalue}) --- see \cite{BSS10a} (section 2 and
Appendix B) for a very detailed discussion. Moreover, quintessence
models usually postulate a preposterously tiny mass for these scalar
fields of the order of $10^{-33}$ eV, i.e. comparable to the current
Hubble rate $H_0$. This is a great disadvantage, not to mention the
fact that they are plagued with the same fine tuning problems as the
original CC approach itself, as emphasized in\,\cite{BSS10a}.

From our point of view, a sensible possibility that should not be
neglected is to think of a ``dynamical $\CC$'' or ``effective vacuum
energy'' $\rLe$, where, rather than replacing $\CC$ by a
collection of \textit{ad hoc} ersatz fields, we stick to the idea
that the CC term in Einstein's equations is still a ``true
cosmological term'', although we permit that ``the observable CC at
each epoch'' can be an effective quantity evolving with the
expansion of the universe: $\rLe=\rLe(H)$, where
\newtext{$H$ is} the Hubble rate. While the
old phenomenological models for time-evolving CC
\,\cite{oldvarCC1,overduin98} lacked of a fundamental motivation, a
general dynamical $\CC$ approach within QFT has more recently been
emphasized in\,\cite{Fossil07,ShapSol09, BAS09,BFLWard09} and also
in the past from the point of view of QFT in curved space-time --
see \,\,\cite{SS12}
and\,\cite{FirstQFTmodels,oldCCstuff1,oldCCstuff2}.

The more modern idea of a slowly running CC is based on the
possibility that quantum effects in curved
space-time\,\cite{Parker09}  can be responsible for the
renormalization group (RG) running of the vacuum energy. One can
show that these models\,\cite{JSP11a} are nicely compatible with the
most recent experimental data\,\cite{BPS09a}, and this fact spurs us
to further explore their impact both in the cosmological and
astrophysical domain. We shall concentrate here on a class of RG
models in which the vacuum energy is a function of cosmic time $t$
though the Hubble rate $H=H(t)$. An archetype example is the quantum
field vacuum model based on the evolution law $\CC(H)=n_0+n_2\,H^2$.
In Ref.\,\cite{BPS09a} some of us have investigated the global
dynamics of this cosmological model (together with various
alternative $\Lambda(t)$ models), in the light of the most recent
cosmological data. However, in that paper we assumed that the vacuum
decayed into matter, whereas here we wish to explore the
complementary possibility that matter is conserved thanks to a
logarithmically variable gravitational coupling $G=G(H)$. This is
very appealing because, if matter is conserved, then one cannot run
into potential problems related with the fact that the vacuum energy
could decay in part into baryons and photons, which would be of
course undesirable. We shall call such FLRW-like model, with
$G=G(H)$ and $\CC=\CC(H)$, the ``running FLRW model'' (hereafter
$\Lambda_{t}G_{t}$CDM), in contrast to the ``running $\Lambda$CDM
model'' (denoted $\Lambda_{t}$CDM) in which only $\CC$ evolves with
time at the expense of maintaining an interaction with
matter\,\cite{oldCCstuff1}.

In this paper, we wish to test the two basic types of running models
in the light of the most recent cosmological data on SNIa+BAO+CMB
(i.e., distant type Ia supernovae, baryonic acoustic oscillations
and cosmic microwave background anisotropies). The clustering
properties of the vacuum energy can also be of high interest as they
may help to shed some light on the fundamental issue of non-linear
structure formation. The subject has been discussed for different DE
models, although mainly based on scalar
fields\,\cite{Lahav91}-\cite{Abramo} (and references therein). A
first study in this direction for running vacuum models was recently
presented by some of us in\,\cite{BPS10a}. In the current work,
however, we emphasize on the implications for the formation and
distribution of the collapsed cosmic structures (i.e. galaxy
clusters) along the lines of the methods utilized in \cite{BPS09a},
which we apply here to the $\Lambda_t G_t$CDM model, and leave the
presentation of the clustering effects in this model for a separate
work. Therefore, we focus here on the study of the dark-matter halo
mass function and the redshift distribution of cluster-size halos,
and compare with the corresponding results for the
$\Lambda_{t}G_{t}$CDM and the concordance $\CC$CDM model.

The paper is organized as follows. In the next section, we discuss
some scenarios with variable $\CC$ and $G$. In Sect.
\ref{sec:runningFLRW}, the FLRW model with running $G$ and $\CC$
(i.e. the $\Lambda_t G_t$CDM model) is introduced, while in Sect.
\ref{sec:runningLCDM} we review the $\Lambda_t$CDM model. The
confrontation of these models with the latest cosmological data is
made in Sect. \ref{sec:ConfrontingRunningFLRW}, and their
predictions concerning the formation of the galaxy cluster-size
halos and the evolution of their abundance is presented in Sect.
\ref{sec:RunningFLRWCollapse}. In the last section, we deliver our
conclusions. Finally, in an appendix we provide details of the
calculation of the linearly extrapolated critical density for the RG
models under consideration.

\section{Models with time-evolving cosmological parameters}\label{VariableParameters}

The idea that the vacuum energy should not be a `rigid' quantity in
cosmology is a most natural one\,\cite{JSP11a}. It is difficult to
conceive an expanding universe with a strictly constant CC energy
density, $\rL=\CC/(8\pi\,G)$, namely one that has remained immutable
since the origin of time. A smoothly-evolving vacuum energy
$\rL=\rL(t)$ that inherits its time-dependence from cosmological
functions such as the scale factor $a(t)$ or the Hubble rate $H(t)$,
is not only a qualitatively more plausible and intuitive idea, but
is also suggested by fundamental physics, in particular by quantum
field theory (QFT) in curved space-time\,\cite{Parker09}. To
implement this notion, it is not strictly necessary to resort to
\emph{ad hoc} scalar fields, as usually done in the literature (e.g.
in quintessence formulations and the like\,\cite{Quintessence}). A
``running'' $\CC$ term can be expected on very similar grounds as
one expects (and observes) the running of couplings and masses with
a physical energy scale in QFT.

Therefore, the guiding paradigm that we adopt in this paper is that
the CC density should naturally be a function of the cosmic time
$t$, or equivalently of the cosmological redshift $z$. The
generalization of this possibility to all the cosmological
parameters, with an eye on trying to explain many surprising
features of the dark energy (DE) -- such as the apparent
phantom-like behavior near our time -- has been put forward in
\cite{SS12}. Within this context, all the cosmological parameters
are viewed as effective couplings whose running can be studied from
the semi-classical formulation of QFT in curved
space-time\,\cite{Parker09}. Interestingly enough, the possibility
to derive this evolution from an effective action has also been
addressed in some particular cases\,\cite{Fossil07,GS11a}. Let us
also remark that these ideas have been pursued in the literature
from different points of view since long ago
\cite{NelPan,FirstQFTmodels,oldCCstuff1,oldCCstuff2,Reuter00}, and
they have been recently re-examined both from the theoretical point
of view\,\cite{ShapSol09} and also from the viewpoint of various
kinds of possible phenomenological
implications\,\cite{BPS09a,GSFS10,BPS10a,SSS04,Letelier10}.

In this framework, the variation of ``fundamental constants'' such
as $\CC$ and $G$ could emerge as an effective description of some
deeper dynamics associated to QFT in curved space-time, or quantum
gravity or even string theory, all of which share the powerful
renormalization group (RG) approach. The latter entails the
possibility to analyze the potential impact of the leading quantum
effects from matter particles on the cosmological observables such
as vacuum energy and gravitational coupling. These effects should
eventually lead to definite time-evolution laws $\rL=\rL(t)\,,
G=G(t)$. However, {given a fundamental model based on QFT}, the
parameter $\rL$ will primarily depend on some cosmological functions
(matter density $\rM$, Hubble expansion rate $H$, etc) which evolve
with time or redshift. Similarly for the Newton coupling $G$.
Therefore, in general we will have the two following functions of
the cosmic time or the cosmological redshift\,\cite{SS12}:
\begin{equation}\label{varibleCCG}
\rL(z)=\rL(\rM(z),H(z),...)\,,\ \ \ \ \ G(z)=G(\rM(z),H(z),...)\,.
\end{equation}
Of course, other fundamental parameters could also be variable. For
example, the fine structure constant has long been speculated as
being potentially variable with the cosmic time (and therefore with
the redshift) including some experimental evidences -- see e.g.
\,\cite{alpharunning}. This might also have an interpretation in
terms of the RG at the level of the cosmological evolution. However,
in this paper we concentrate purely on the potential variability of
the most genuinely fundamental gravitational parameters of
Einstein's equations such as $\CC$ and $G$. The possibility that $G$
could be variable has been previously entertained in the literature,
although most of these formulations are usually framed within the
original Jordan and Brans-Dicke proposals\,\cite{JBD} linked to the
existence of a dynamical scalar field coupled to curvature.

The cosmological constant contribution to the curvature of
space-time is represented by the $\Lambda\,g_{\mu\nu}$ term on the
\textit{l.h.s.} of Einstein's equations, which can be absorbed on
the \textit{r.h.s.} of these equations:
\begin{equation} R_{\mu \nu
}-\frac{1}{2}g_{\mu \nu }R=8\pi G\ \tilde{T}_{\mu\nu}\,, \label{EE}
\end{equation}
where the modified energy-momentum tensor is given by
$\tilde{T}_{\mu\nu}\equiv T_{\mu\nu}+g_{\mu\nu}\,\rL $. Here
$\rL=\CC/(8\pi G)$ is the vacuum energy density associated to the
presence of $\CC$, and $T_{\mu\nu}$ is the ordinary energy-momentum
tensor of isotropic matter and radiation. Modeling the expanding
universe as a perfect fluid with velocity $4$-vector field
$U^{\mu}$, we have
$T_{\mu\nu}=-\pM\,g_{\mu\nu}+(\rM+\pM)U_{\mu}U_{\nu}$, where $\rM$
is the proper isotropic density of matter-radiation and $\pM$ is
the corresponding pressure. Clearly the modified
$\tilde{T}_{\mu\nu}$ defined above takes the same form as
${T}_{\mu\nu}$ with $\tilde{\rho}=\rM+\rL$ and $\tilde{p}=\pM-\rL$.
Therefore:
\begin{equation}
\tilde{T}_{\mu\nu}=-\tilde{p}\,g_{\mu\nu}+(\tilde{\rho}+\tilde{p})U_{\mu}U_{\nu}=
(\rL-\pM)\,g_{\mu\nu}+(\rM+\pM)U_{\mu}U_{\nu}\,. \label{Tmunuideal}
\end{equation}
With this generalized energy-momentum tensor, and in the spatially
flat FLRW metric  $ds^{2}=dt^{2}-a^{2}(t)d\vec{x}^{2}$, the
gravitational field equations boil down to Friedmann's equation
\begin{equation}
H^{2}\equiv \left( \frac{\dot{a}}{a}\right) ^{2}=\frac{8\pi\,G }{3}%
\tilde{\rho}=\frac{8\pi\,G }{3}%
\left( \rM +\rL\right)\,,  \label{FL1}
\end{equation}
and the dynamical field equation for the scale factor:
\begin{equation}
\ddot{a}=-\frac{4\pi}{3}G\,(\tilde{\rho}+3\,\tilde{p})\,a=-\frac
{4\pi}{3}G\,(\rM+3\,\pM-2\,\rL)\,a\,. \label{newforce3}
\end{equation}
Let us next contemplate the possibility that $G=G(t)$ and
$\rL=\rL(t)$ can be both functions of the cosmic time. This is
allowed by the Cosmological Principle embodied in the FLRW metric.
The Bianchi identities (which insure the covariance of the theory)
then imply
$\bigtriangledown^{\mu}\,\left({G\,\tilde{T}}_{\mu\nu}\right)=0$.
In the case of the FLRW metric, the previous identity amounts to
the following generalized local conservation law:
\begin{equation}\label{BianchiGeneral}
\frac{d}{dt}\,\left[G(\rL+\rM)\right]+3\,G\,H\,(\rM+\pM)=0\,.
\end{equation}
This equation is actually a first integral of the dynamical system
(\ref{FL1}) and (\ref{newforce3}), as can be easily checked. When
$G$ is constant, the identity above implies that $\rL$ is also a
constant, if and only if the ordinary energy-momentum tensor is
individually conserved ($\bigtriangledown^{\mu}\,{T}_{\mu\nu}=0$),
i.e. $\dot{\rho}_m+3\,H\,(\rM+\pM)=0$. This equation can be
rewritten
\begin{equation}\label{mattercons}
{\rho}_m'(a)+\frac3a\,(\rM(a)+\pM(a))=0\,,
\end{equation}
where the prime indicates differentiation with respect to the scale
factor: $f'=df/da$ for any function $f(a)$. The solution of
(\ref{mattercons}) is
\begin{equation}\label{solstandardconserv}
\rM(a)=\rM^0\,a^{-3(1+\omm)}=\rM^0\,(1+z)^{3(1+\omm)}
%\rM(a)=\rM^0\,a^{-n_m}=\rM^0\,(1+z)^{n_m}\,,\ \ \
%n_m=3(1+\omm)\,,
\end{equation}
where $\rM^0$ is the current matter density and $\omm=\pM/\rM$ is
the equation of state (EoS) parameter for cold ($\omm=0$) and
relativistic ($\omm=1/3$) matter, respectively. We have expressed
the result (\ref{solstandardconserv}) in terms of the scale factor
$a=a(t)$ and the cosmological redshift $z=(1-a)/a$.

A first non-trivial situation appears when $\rL=\rL(t)$ but $G$
remains still constant. Then equation (\ref{BianchiGeneral}) boils
down to
\begin{equation}\label{Bronstein}
\dot{\rho}_{\CC}+\dot{\rho}_m+3\,H\,(\rM+\pM)=0\,,
\end{equation}
This scenario shows that a time-variable $\rL$ cosmology may exist
such that transfer of energy occurs from matter-radiation into
vacuum energy, and vice versa (see section 4). However, let us
remark that the time evolution of $\rL$ is still possible in a
framework where matter is strictly conserved, i.e. such that
equation (\ref{mattercons}) is maintained. Such scenario is highly
desirable and is perfectly consistent with the Bianchi identity at
the expense of a time-varying gravitational coupling: $G=G(t)$.
Indeed, it is easy to see that equation (\ref{mattercons}) is
compatible with (\ref{BianchiGeneral}) provided that the following
differential constraint between the variable $G$ and the variable
$\rL$ is satisfied:
\begin{equation}\label{Bianchi}
(\rM+\rL)\,d{G}+G\,d{\rho}_{\CC}=0\,,
\end{equation}
where $\rM$ in this equation is given by (\ref{solstandardconserv}).
To solve the model in the basic set of variables $(H(z),\rM(z),p(z),
G(z), \rL(z))$, still another equation is needed. In the next
section we discuss the possibility that the fifth missing equation
is a QFT-inspired relation of the form $\rL=\rL(H(z))$.

\section{A FLRW-like model with running $G$ and $\CC$:
the $\Lambda_{t}G_{t}$CDM model}\label{sec:runningFLRW} Consider the
class of models in which the vacuum energy and the gravitational
coupling evolve as power series of some energy scale $\mu$, such
that the rates of change are given respectively by
\begin{equation}\label{seriesRLG}
\frac{d\rL(\mu)}{d\ln\mu}= \sum_{k=0,1,2,...}\,A_{2k}\,\mu^{2k}\,,\
\ \ \ \ \ \frac{d}{d\ln\mu}\left(\frac{1}{G(\mu)}\right)=
\sum_{k=0,1,2,...}\,B_{2k}\,\mu^{2k}\,.
\end{equation}
These expansions can be considered as purely phenomenological
ansatzs or, if we aim at a more fundamental description, as emerging
from the quantum field RG running of the cosmological
parameters\,\cite{FirstQFTmodels,oldCCstuff1,SSS04}. Such running
should ultimately reflect the dependence of the leading quantum
effects on some physical cosmological quantity $\xi$ associated with
$\mu$, hence $\rL=\rL(\xi)$. The physical scale $\xi$  could
typically be the Hubble rate $H$, or even the scale factor $a$,
which in most of the cosmological past also maps out the evolution
of the energy densities with $H$\,\cite{Fossil07}. Since in both
cases $\xi=\xi(t)$ evolves with the cosmic time, the cosmological
term inherits a time-dependence through its primary scale evolution
with $H(t)$ or $a(t)$. In this context, the \textit{r.h.s.} of the
above equations defines the $\beta$-functions for the running of
$\rL$ and $G^{-1}$ in QFT in curved space-time. The setting $\mu\to
H$ naturally points to the typical energy scale of the classical
gravitational external field associated to the FLRW metric, i.e. the
characteristic energy of the FLRW ``gravitons'' attached to the
quantum matter loops contributing to the running of $\rL$ and
$G^{-1}$ in a semi-classical description of gravity. Coefficients
$A_{2k}, B_{2k}$ in these formulas receive contributions from boson
and fermion matter fields of different masses $M_i$, and the series
(\ref{seriesRLG}) become expansions in powers of the small
quantities $H/M_i$\,\cite{oldCCstuff1,oldCCstuff2}. The fact that
only even powers of $H$ are involved is dictated by the general
covariance of the effective
action\,\cite{ShapSol09,Fossil07}\,\footnote{In practice, if one
tries to fit the data with a time dependent CC term which is linear
in the expansion rate, i.e. of the form $\CC\propto H$, the results
deviate significantly from the standard $\CC$CDM
predictions\,\cite{BPS09a,LinearH}.}. After integrating the formulas
(\ref{seriesRLG}), the result may be looked upon as a specific
implementation of the general equations (\ref{varibleCCG}), where
the expansion rate $H$ is singled out as the main cosmological
variable on which the functions $\rL$ and $G^{-1}$ depend and in
terms of which are expanded.

It is important to note that the expansions (\ref{seriesRLG}) are
correlated by the Bianchi identity (\ref{Bianchi}) and therefore
they must be consistent with it, order by order. Let us consider
them more closely. We start by first rewriting the expression for
the RG of the CC term in a more explicit manner, as follows:
\begin{equation}\label{seriesLambda}
\frac{d\rL(\mu)}{d\ln\mu}=\frac{1}{(4\pi)^2}\left[\sum_{i}\,B_{i}M_{i}^{2}\,\mu^{2}
+\sum_{i}
\,C_{i}\,\mu^{4}+\sum_{i}\frac{\,D_{i}}{M_{i}^{2}}\,\mu^{6}\,\,+...\right]
\equiv 2\,n_2\,\mu^2+{\cal O}(\mu^4)\,,
\end{equation}
where $M_{i}$ are the masses of the particles contributing in the
loops, and $B_{i},C_i,..$ are dimensionless parameters. We have
omitted the $M_i^4$ terms in the expansion since, as explained in
\cite{oldCCstuff1,oldCCstuff2}, such terms would trigger a too fast
running of the CC.  It follows that only the ``soft-decoupling''
terms of the form $\sim M_i^2\,\mu^2$ remain in practice. The
integrated form of (\ref{seriesLambda}) reads
\begin{equation}\label{GeneralPS}
\rL(H)=n_{0}+n_{2}H^{2}\,,
\end{equation}
where we have set $\mu=H$ and neglected the smaller higher order
terms, which are of ${\cal O}(H^4)$. The additive constant $n_0$ in
(\ref{GeneralPS}) must be essentially given by the current value of
the vacuum energy, $\rLo$, but not quite. Notice that if $n_2=0$,
the CC is strictly constant and then $n_0$ just coincides with
$\rLo$. However, if the $H^2$ term is to play a significant role
while preserving the approximate $\CC$CDM behavior, it should
neither be negligible nor dominant as compared to $n_0$. Being the
latter the leading term in the series expansion, it must still be of
order $n_0\sim\rLo$; thus the largest masses $M_i$  associated to
$n_2\sim\sum_i B_i\,M^2_i$ should be in the ballpark of the mass
$M_X$ associated to a typical Grand Unified Theory (GUT) near the
Planck scale $M_P$ (see below). From the structure of the expansion
(\ref{seriesLambda}), with $\mu=H$ and $M_i\sim M_X$, we reconfirm
that no other even power $H^{2n}$ (not even $H^4$) can contribute
significantly on the \textit{r.h.s.} of equation (\ref{GeneralPS})
at any stage of the cosmological history below $M_X\lesssim M_P$. It
is particularly convenient to rewrite the coefficients of
(\ref{GeneralPS}) as follows:
\begin{equation}\label{n0n2}
n_0=\rLo-\frac{3\nu}{8\pi}\,M_P^2\,H_0^2\,,\ \ \ \ \
n_2=\frac{3\nu}{8\pi}\,M_P^2\,,
\end{equation}
with
\begin{equation}\label{nu1}
\nu=\frac{1}{6\pi}\, \sum_i B_i\frac{M_i^2}{M_P^2}\,.
\end{equation}
Here, $H_0$ is the value of the Hubble rate at present. A most
important (dimensionless) parameter of the present framework is
$\nu$, given in equation (\ref{nu1}). It defines the main
coefficient of the $\beta$-function for the running of the vacuum
energy\,\cite{oldCCstuff1}. For $\nu=0$, the vacuum energy remains
strictly constant, $\rL=\rLo$, whereas for non-vanishing $\nu$  the
evolution law (\ref{GeneralPS}) can now be written as\,\footnote{It
is very interesting to notice that this quadratic evolution law for
the vacuum energy with the expansion rate has also been suggested
recently by alternative QFT methods, see \cite{Maggiore10}.}
\begin{equation}\label{RGlaw2}
 \rL(H)=\rLo+ \frac{3\nu}{8\pi}\,M_P^2\,(H^{2}-H_0^2)\,.
\end{equation}
Let us stress that, in the QFT framework, $\nu$ is naturally
expected to be non-vanishing owing to the quantum effects of matter
particles. The coefficients $B_i$ in (\ref{nu1}) can be computed
from the quantum loop contributions of fields with masses $M_i$, see
e.g. \cite{Fossil07} for a specific framework. It is customary to
write $\nu$  in the compact form
\begin{equation}\label{nu}
\nu=\frac{\sigma}{12\pi}\,\frac{M^2}{M_P^2}\,,
\end{equation}
with $M$  an effective mass parameter of order of the average mass
of the heavy particles of the underlying GUT, including their
multiplicities \,\cite{oldCCstuff1}. From the expression
(\ref{nu1}), it is clear that the natural range of that parameter is
$|\nu|\ll 1$; and it can be both positive or negative, since
$\sigma=\pm 1$ depending on whether bosons or fermions dominate in
the sum of loop contributions (\ref{nu1}).  For instance, if GUT
fields with masses $M_i$ near $M_P$ do contribute, then
$|\nu|\lesssim 1/(12\pi)\simeq 2.6\times 10^{-2}$, but in general we
expect it to be smaller because the usual GUT scales, say $M_X\sim
10^{16}$ GeV, are not that close to $M_P\sim 10^{19}$ GeV. By
counting heavy particle multiplicities in a typical GUT (i.e. the
total number of particles with masses $M_i\sim M_X$), a natural
estimate lies in the range $\nu=10^{-5}-10^{-3}$ \,\cite{Fossil07}.

Clearly, the vacuum energy density (\ref{RGlaw2}) is normalized to
the present value, i.e.
\begin{equation}\label{rLo}
\rL(H_0)=\rLo\equiv\frac{3}{8\pi}\,\OLo\,H^{2}_{0}\,M_P^2\,.
\end{equation}
Recall that the observed numerical value of this expression is given
by equation (\ref{CCvalue}), corresponding to $\OLo\simeq 0.7$. We
remark that the parametrization (\ref{RGlaw2}) satisfies the
aforementioned condition that $n_0$ is the leading term and is of
order $\rLo$, and at the same time the correction term is of order
$M^2\,H^2$, with $M\sim \sqrt{\nu}\,M_P$ a large mass, even if $\nu$
is as small as, say, $|\nu|\sim 10^{-3}$ or less.

What about the running equation for $G^{-1}$ in equation
(\ref{seriesRLG})? A more detailed form for it reads:
\begin{equation}\label{RGEG}
\frac{d}{d\ln \mu}\left(\frac{1}{G}\right)=
\sum_i\,a_i\,M_i^2+\sum_i\,b_i \mu^2+
\sum_i\,c_i\,\frac{\mu^4}{M_i^2}+...
\end{equation}
If we take again $\mu=H$, then, since we have $H\ll M_i$ for all the
relevant epochs of the universe -- at and below the GUT scale --,
this series should converge extremely fast, the first term being the
dominant one and the remaining terms are inessential. The first term
is of course a contribution to the running of $G^{-1}$ which is
directly driven by the heavy particle masses $M_i$ of the GUT, and
therefore can be parametrized as a coefficient times the Planck
mass. Thus, we are essentially led to a RGE of the type
\begin{equation}\label{RGEG2}
\frac{d}{d\ln \mu}\left(\frac{1}{G}\right)= 2\,\nu\,M_P^2+...\,,
\end{equation}
where the coefficient in front of $M_P$ has been fixed by the
condition that this solution is compatible with the Bianchi identity
(\ref{Bianchi}) after we adopt the solution (\ref{RGlaw2}) for the
evolution of the vacuum energy and set $\mu=H$, too, for the running
$G$ (see below). We shall show below explicitly this consistency,
but let us first notice that the solution of (\ref{RGEG2})
satisfying the boundary condition $G(\mu=H_0)=G_0\equiv 1/M_P^2$ is
\begin{equation}\label{GH} g(H)\equiv
\frac{G(H)}{G_0}=\frac{1}{1+\nu\,\ln\left(H^2/H_0^2\right)}\,.
\end{equation}
We see that $g(H)$ depends on the parameter $\nu$ and that, to first
order, we have $g(H)=1+{\cal O}(\nu)$, i.e. $G(H)=G_0(1 +{\cal
O}(\nu))$. Thus, $\nu$ plays also (in a very evident manner here)
the role of the $\beta$-function for the RG running of $G$. For
example, for $\nu>0$ the gravitational coupling decreases with the
characteristic cosmological energy $H$ (hence $G$ increases with the
expansion); it follows that for $\nu>0$ the coupling $G$ is
asymptotically free (as the QCD coupling at high energies), whereas
for $\nu<0$ the gravitational coupling behaves as the
electromagnetic one, i.e. increases with the energy (equivalently,
$G$ decreases in this case with the expansion). In both cases the
running of $G$ is extremely slow, since it is logarithmic and with a
very small coefficient $|\nu|\ll 1$. Let us note that if we would
have omitted the leading $\sim M_i^2$ term on the \textit{r.h.s.} of
equation (\ref{RGEG}) -- as we have done with the $M_i^4$ term on
the \textit{r.h.s.} of equation (\ref{seriesLambda}) -- we would
have obtained a still slower running of the gravitational coupling:
\begin{equation}\label{xirunning}
g(H)=\frac{1}{1+\xi\,\,(H^2-H_0^2)/M_P^2}\,,
\end{equation}
where $\xi$ is a dimensionless coefficient. In deriving equation
(\ref{xirunning}) we have assumed that the \textit{r.h.s.} of
(\ref{RGEG}) is dominated by the next-to-leading term $\sim H^2$
(for $\mu=H$) under the assumption that the $\sim M_i^2$ terms
cancel. Since $H^2/M_P^2$ is utterly negligible, this new running
law for $G$ is even less sensitive to time evolution for the present
universe. In this case, the corresponding evolution of the CC
following from the Bianchi identity (\ref{Bianchi}) would also be
very mild:
\begin{equation}\label{CCxirunning}
\rL(H)=\rLo+\frac{3\xi}{16\,\pi}\,\left(H^4-H_0^4\right)\,.
\end{equation}
However, since there is no a priori theoretical reason nor a
phenomenological constraint to assume that the leading term $\sim
M_i^2$ terms cancel, we shall concentrate here only on the form
(\ref{GH}) and corresponding (\ref{RGlaw2}), rather than on
(\ref{xirunning})-(\ref{CCxirunning}), when we confront our running
model with the observational data in section
\ref{sec:ConfrontingRunningFLRW}.

Before checking the consistency of the solution (\ref{GH}), it is
convenient to introduce some notation. We define the energy
densities normalized with respect to the \emph{current} critical
density $\rco=3H_0^2/(8\pi\,G_0)$:
\begin{eqnarray}
\Omega_i(z)&\equiv&\frac{\rho_i(z)}{\rco} \ \ \
(i=m,\CC)\,.\label{Omegas}
\end{eqnarray}
It is also convenient to introduce the energy densities normalized
with respect to the critical density at an arbitrary redshift,
$\rc(z)=3H^2(z)/(8\pi\,G(z))$, in which both $H$ and $G$ are
functions of $z$. This new set of normalized energy densities is
given by
\begin{eqnarray}
\tilde{\Omega}_i(z)&\equiv&\frac{\rho_i(z)}{\rc(z)}=\frac{g(z)}{E^2(z)}\,\Omega_i(z)\
\ \ \ (i=m,\CC)\,,\label{Omegastilde}
\end{eqnarray}
where we have defined the normalized Hubble rate of the running FLRW
model with respect to the current one:
\begin{equation}\label{Eg}
E(z)\equiv\frac{H(z)}{H_0}=\sqrt{g(z)}\,\left[\OM(z)+\OL(z)\right]^{1/2}\,,
\ \ \ g(z)\equiv\frac{G(z)}{G_0}\,.
\end{equation}
This is the generalized Friedmann's equation for the present model.
Obviously, the parameters (\ref{Omegastilde}) coincide with the
previous ones (\ref{Omegas}) only at $z=0$, and at that point they
all furnish the normalized current densities:
$\Omega_i(0)=\tilde{\Omega}_i(0)=\Omega_i^0$. However, only the
tilded cosmological parameters (\ref{Omegastilde}) satisfy the (flat
space) cosmic sum rule (valid at any redshift $z$):
\begin{equation}\label{sumrule}
\tilde{\Omega}_m(z)+\tilde{\Omega}_{\CC}(z)=1\,.
\end{equation}
We agreed to call the FLRW-like model with variable $G=G(H)$ and
$\CC=\CC(H)$ the $\Lambda_{t}G_{t}$CDM model, in contrast to the
$\Lambda_{t}$CDM model (previously studied in \cite{oldCCstuff1} and
most recently in great detail in \cite{BPS09a}), in which only $\CC$
evolves with $H$ at the expense of maintaining an interaction with
matter through the generalized conservation law (\ref{Bronstein}) --
see the next section for a summarized discussion of the
$\Lambda_{t}$CDM model.

The solution of the running FLRW model is determined from the
Friedmann equation (\ref{Eg}), the Bianchi identity (\ref{Bianchi})
and the RG law for the CC (\ref{RGlaw2}). Using the normalized
density parameters defined in equation (\ref{Omegas}), the basic
cosmological equations can be formulated in the following compact
way:
\begin{eqnarray}\label{System1}
&&E^2(z)=g(z)\left[\OM(z)+\OL(z)\right]\label{Friedmann2}\,,\\
&&(\OM+\OL)dg+g\,d\OL=0\,,\label{bianchi2}\\
&& \OL(z)=\OLo+\nu\left[E^2(z)-1\right]\,, \label{RGlaw3}\\
%&&\OM(z)=\OM^0\,(1+z)^{3(1+\omega_m)}\,,\label{conservOm}
&&\OM(z)=\OM^0\,(1+z)^{3(1+\omm)}\,,\label{conservOm}
\end{eqnarray}
where the last equation just reflects the covariant conservation of
matter (for cold dark matter we have $\omm=0$ and for relativistic
matter we use $\omm=1/3$), i.e. it is a rephrasing of equation
(\ref{solstandardconserv}). Solving the differential form
(\ref{bianchi2}) in combination with the rest of the equations, it
is easy to arrive at
\begin{equation}\label{System2}
\frac{dg}{g^2}=-\frac{d\OL}{E^2}=-2\nu\,\frac{dE}{E}\,.
\end{equation}
Integrating this expression with the boundary condition $g(z=0)=1$,
and using also the fact that $E(z=0)=1$, we finally meet the
explicit form (\ref{GH}) for the function $g=g(H)$, in full
consistency with our expectations --- \textit{q.e.d.}.

We may also check a posteriori the consistency of the scale choice
$\mu=H$ for the running of the cosmological parameters. The election
of this scale is of course a difficult issue in cosmology, as there
are different possibilities at our disposal --
see\,\cite{oldCCstuff1,Fossil07} and references therein. However, we
expect that the choice $\mu=H$ should naturally be self-consistent
in the FLRW context at the leading order. We can substantiate this
assertion starting from the generic functions (\ref{varibleCCG}),
which we assume that are dominated by a single scaling variable
$\mu$, i.e. $\rL=\rL(\mu)$ and $G=G(\mu)$. The differentials of
these variables in the Bianchi identity (\ref{Bianchi}) can then be
written in terms of $d\mu$. We easily arrive at the following
expression:
\begin{equation}\label{BianchiH}
G(\mu)\left(\rM+\rL(\mu)\right)=\frac{d\rL}{d\ln
\mu}\left(\frac{d}{d\ln
\mu}\,\frac{1}{G(\mu)}\right)^{-1}=\frac{3}{8\pi}\,\mu^2\,.
\end{equation}
where in order to evaluate the final form on its \textit{r.h.s.} we
have used the running laws (\ref{seriesLambda}) and (\ref{RGEG2}) as
well as the explicit form of the coefficient $n_2$ in (\ref{n0n2}).
Notice that the small parameter $\nu$ has neatly canceled on the
\textit{r.h.s.} of (\ref{BianchiH}), within the order under
consideration. At the end of the day, equation (\ref{BianchiH})
tells us that $\mu$ is determined by the simple formula
\begin{equation}\label{muequalH}
\mu^2=\frac{8\pi\,G(\mu)}{3}\left[\rM+\rL(\mu)\right]\,,
\end{equation}
which patently shows, upon comparison with (\ref{FL1}), that $\mu=H$
for the flat space
--- \textit{q.e.d.} Therefore, the identification $\mu=H$ is
consistent within the order under consideration. Furthermore, it
also suggests that the evolution laws (\ref{RGlaw2}) and (\ref{GH})
can be interpreted within the context of the RG in cosmology, with
$\mu=H$ acting as the natural running scale.

Let us now come back to the running of the gravitational coupling
with $\mu=H$. Although the form (\ref{GH}) is the simplest and most
revealing one from the physical point of view, it is also convenient
to determine the dependence of the gravitational coupling as a
function of the cosmological redshift, i.e. $g=g(z)$. This function
is necessary for the numerical analysis of the model and the
comparison with the observational data. To this end, we start by
combining equations (\ref{Friedmann2}) and (\ref{RGlaw3}), and we
obtain:
\begin{equation}\label{OLz}
\OL(z)=\frac{\OLo+\nu\left[\OM(z)\,g(z)-1\right]}{1-\nu\,\,g(z)}\,.
\end{equation}
From the above equations we also have
\begin{equation}\label{OMplusOL}
\OM(z)+\OL(z)=\frac{\OM(z)+\OLo-\nu}{1-\nu g(z)}\,.
\end{equation}
On comparing it with the sum rule (\ref{sumrule}) for the tilded
parameters, we see that only for $z=0$ the \textit{r.h.s.} of
equation (\ref{OMplusOL}) gives $1$, as expected. Next we compute
$d\OL(z)$ explicitly from equation (\ref{OLz}) and substitute the
obtained result, together with equation (\ref{OMplusOL}), in the
differential form (\ref{bianchi2}). After some straightforward
algebra, we find
\begin{equation}\label{diffform2}
(\OM(z)+\OLo-\nu)\,dg+\nu\,(1-\nu\,g)\,g^2\,d\OM(z)=0\,,
\end{equation}
which can be integrated by quadrature:
\begin{equation}\label{implicitgz}
\frac{1}{g(z)}-1+\nu\,\ln\left[\frac{1}{g(z)}-\nu\right]=
\nu\,\ln\left[\OM(z)+\OLo-\nu\right]\,.
\end{equation}
Notice that $\OM(z)$ in the previous equation is given by
(\ref{conservOm}), but in those cases when there is a comparable
mixture of cold matter and radiation we have to write
\begin{equation}\label{mixture}
\OM(z)=\OM^0\,(1+z)^{3}+\OR^0\,(1+z)^{4}\,,
\end{equation}
where $\ORo=(1+0.227 N_{\nu})\, \Omega_{\gamma}^0$, with $N_{\nu}$
the number of neutrino species and $\Omega_{\gamma}^0\,h^2\simeq
2.47\times 10^{-5}$. This is the case, for example, when we analyze
the matter content near the decoupling time for the analysis of the
CMB observables, see the next section.

Equation (\ref{implicitgz}) determines $g=g(z)$ as an implicit
function of the redshift. There is no obvious way to write the
explicit function. It is nevertheless easy to check that it
satisfies the boundary condition $g(0)=1$, as it should. With $g(z)$
determined from (\ref{implicitgz}), the redshift evolution of the
vacuum energy follows from (\ref{OLz}) and (\ref{mixture}). For the
present model, however, it is not possible to find an analytical
expression of the cosmological functions with respect to the cosmic
time, only in terms of the redshift, or equivalently the scale
factor. But this is enough for the numerical analysis.

As we know, the coefficient $\nu$ in equation\,(\ref{RGlaw2}), or
equivalently (\ref{RGlaw3}), measures the amount of running of the
CC or vacuum energy in the running FLRW model. For any given $\nu$,
we can compare the value of $\OL(z)$ with the current value
$\OLo\simeq 0.7$. The relative correction can be conveniently
expressed as follows:
\begin{equation}\label{DeltarLo}
\Delta\OL(z)\equiv\frac{\OL(z)-\OLo}{\OLo}=\frac{\nu}{\OLo}\left[E^2(z)-1\right]\,.
\end{equation}
Since $g=1$ for $\nu=0$, it follows that, for small $\nu$, $g(z)$
deviates little from $1$, namely $g(z)=1+{\cal O}(\nu)$. Thus,
expanding to order $\nu$ in the matter epoch, it is easy to show
from the previous equation that
\begin{equation}\label{deltarhoCC}
\Delta\OL(z)\simeq
\nu\,\frac{\OM^0}{\OLo}\,\left[(1+z)^3-1\right]\,,
\end{equation}
where $g(z)\sim 1$ to this order. If we look back to relatively
recent past epochs, e.g. exploring redshifts $z={\cal O}(1)$
relevant for Type Ia supernovae measurements, the deviation
$\Delta\OL(z)$ is of order of a few times $\,\nu$. For example, for
$z=1.5$ and $z=2$, the deviations are $\Delta\OL(1.5)\simeq 6\,\nu$
and $\Delta\OL(2)\simeq 11\,\nu$ respectively, assuming $\OM^0=0.3$.
Although the correction is small, it is not necessarily negligible,
even if $|\nu|\sim 10^{-3}$. In the next sections, we will see that
the analysis of the cosmological observables in the light of the
latest SNIa+BAO+CMB data amounts to a bound on $|\nu|$ of this
order, and therefore there is hope to eventually becoming sensitive
to the effects of the running vacuum energy.

Since we will also use the data on structure formation in order to
constraint the running $\Lambda_{t}G_{t}$CDM model in section
\ref{sec:ConfrontingRunningFLRW}, let us recall that in any model
with variable G and $\rL$ in which matter is covariantly conserved,
the matter density contrast $D\equiv {\delta\rho_m}/{\rho_m}$
satisfies the following second order differential equation with
respect to the scale factor\,\cite{GSFS10}:
\begin{eqnarray}
D''(a)+\left(\frac{3}{a}+\frac{H'(a)}{H(a)}\right)D'(a)=\frac{3\tilde{\Omega}_m(a)}{2a^2}\left(D(a)+\frac{\delta
G}{G}\right)\,,\label{difff1}
\end{eqnarray}
where the cosmological parameter $\tilde{\Omega}_m(a)$ has been
defined in (\ref{Omegastilde}). Moreover, the perturbations in $G$
are actually tightly linked to those in $\rL$, and they are at the
same time linked to the matter perturbations, as follows:
\begin{equation}\label{correlation}
\frac{\delta\rho_{\Lambda}}{\rho_{\Lambda}}=-\frac{\delta G}{G}\,, \
\ \ \ \ \ \ \ \ \frac{\delta\rho_m}{\rho_m}=-\frac{{(\delta
G(a))'}}{{G'(a)}}\,.
\end{equation}
If we neglect the perturbations in $G$ ($\delta G\sim0$) and hence
in $\rL$ too, equation (\ref{difff1}) reduces to the standard form
of the growth factor for models with self-conserved matter (under
the assumption of negligible DE perturbations), as for example in
the case of matter perturbations in the $\CC$CDM
model\,\cite{Peeb93}. If, on the contrary, we consider non-vanishing
$\delta G$, and hence non-vanishing $\delta\rL$, one arrives at the
following third order differential equation:
\begin{equation}
D'''(a)+\frac{1}{2}\left(16-9\tilde{\Omega}_m(a)\right)\frac{D''(a)}{a}+\frac{3}{2}\left(8-11\tilde{
\Omega}_m(a)+3\tilde{\Omega}_m^2(a)-a\,\tilde{\Omega}_m'(a)\right)\frac{D'(a)}{a^2}=0\,.\label{supeq}
\end{equation}
This equation was derived analytically (in both the synchronous and
Newtonian gauges) in\,\cite{GSFS10}, where it was also numerically
analyzed.
%In the present
%paper, however, in order to better compare with the previous results
%for the $\CC_t$CDM model\,\cite{BPS09a}, we will limit ourselves to
%consider the case with no perturbations in $G$, and thus in
Let us stress that when $G$ is variable, it is necessary to include
the perturbations in this variable since, as we have seen, they
become correlated with the perturbations in $\rL$ because of the
Bianchi identity of the running FLRW model, equation
(\ref{Bianchi}). Moreover, there is no consistent way to set them to
zero if the matter perturbations themselves are nonzero, see
equation (\ref{correlation}). Therefore, a fully consistent
treatment of the perturbations for this model requires to take into
account perturbations on all these variables at the same time. This
is done automatically by the third order differential equation
(\ref{supeq}). In sect.\,\ref{sec:ConfrontingRunningFLRW} we will
use this equation for the study of the structure formation
properties within the running FLRW model.

\section{A FLRW-like model in which only $\CC$ varies:
The $\Lambda_{t}$CDM model}\label{sec:runningLCDM}

Let us now consider the possibility of interacting
$\Lambda$-cosmology, that is, a cosmology in which the vacuum
exchanges energy with matter at a fixed value of $G=G_0$. The global
dynamics of such models has been investigated extensively in the
literature on purely phenomenological grounds, in fact, much before
the discovery of the present accelerating stage \cite{oldvarCC1} --
see\,\cite{overduin98} for a review of the old models. However, a
variant of these models which can be well motivated within the
framework of QFT in curved space-time did not appear until
later\,\cite{FirstQFTmodels} and in subsequent
works\,\cite{oldCCstuff1,oldCCstuff2}. Here we briefly present the
main points of the theoretical analysis -- for an updated and
thorough discussion, see Basilakos et al. \cite{BPS09a}. Following
our general QFT framework of section \ref{sec:runningFLRW}, we
expect a functional dependence with $H$, i.e. $\rL=\rL(H)$, in which
the first non-trivial power with respect to the Hubble rate is
$H^2$, meaning that the basic dependence with $H$ is again of the
form (\ref{GeneralPS}), or equivalently (\ref{RGlaw2}). Since $G$ is
constant now, the exchange of energy between vacuum and matter is
governed by equation (\ref{Bronstein}). Combining the latter with
Friedmann's (\ref{FL1}) for a flat universe and the acceleration law
(\ref{newforce3}), one obtains
\begin{equation}
 \label{frie34bis}
\dot{H}+\frac{3}{2} H^{2}=4\pi\,G\,\left(\rL-\omm\rmm\right)\,,
\end{equation}
where $\omm=p_m/\rmm$. In the particular case $\omm=0$ (matter
dominated epoch), we have\,\cite{BPS09a}
\begin{equation}
 \label{frie34}
\dot{H}+\frac{3}{2} H^{2}=\frac{\Lambda}{2}=4\pi\,G\,\rL\,.
\end{equation}
The traditional cosmology (i.e. the concordance $\CC$CDM model) can
be obtained as a particular case of this equation by direct
integration of it for $\rL=$const., but the same equation
(\ref{frie34}) is also valid for $\rL=\rL(H)$. Therefore, if we
insert (\ref{RGlaw2}) in (\ref{frie34}) we can obtain the
corresponding Hubble expansion rate for the $\Lambda_{t}$CDM model.
It turns out that, in this case, an explicit integration with
respect to the cosmic time is feasible (a feature which was
impossible for the $\Lambda_{t}G_{t}$CDM model of section
\ref{sec:runningFLRW}). The final result is\,\cite{BPS09a}
\begin{equation}
\label{frie455} H(t)=H_{0}\,\sqrt{\frac{\OLo-\nu}{1-\nu}} \;
\coth\left[\frac32\,H_{0}\sqrt{(\OLo-\nu)(1-\nu)}\;t\right]\,.
\end{equation}
From here the corresponding scale factor $a(t)$ can also be found
immediately:
\begin{equation}\label{frie456}
a(t)=\left(\frac{\OMo}{\OLo-\nu}\right)^{\frac{1}{3(1-\nu)}}\,
\sinh^{\frac{2}{3(1-\nu)}}
\left[\frac32\,H_{0}\sqrt{(\OLo-\nu)(1-\nu)}\;t\right]\,.
\end{equation}
Eliminating the time variable between equations (\ref{frie455}) and
(\ref{frie456}) we arrive at an expression of the normalized Hubble
flow in terms of the scale factor, or equivalently the redshift
$z=(1-a)/a$:
\begin{equation}
E(z)=\frac{H(z)}{H_0}=
\left[\frac{\OLo-\nu}{1-\nu}+\frac{\Omega^{0}_{m}}
{1-\nu}(1+z)^{3\,(1-\nu)}\right]^{1/2} \;. \label{nomalHflow}
\end{equation}
Let us finally report on the behavior of the matter and vacuum
energy densities in this model as a function of the redshift. This
is an important point as it shows one of the basic differences
between this model and the one considered in the previous section.
At this point we will come back to the general matter EoS parameter
$\omm=\pM/\rM$ because we wish to see the change in the evolution of
both cold DM and relativistic particles. Starting from the covariant
conservation law (\ref{Bronstein}), and trading the time derivatives
for derivatives with respect to the redshift (through the relation
$d/dt=-H(1+z)d/dz$), we obtain
\begin{equation}\label{Bronstein2}
\frac{d\rmm}{dz}-\,\frac{3(1+\omm)\rmm}{1+z} =-\,\frac{d\rL}{dz}\,.
\end{equation}
Using this equation in combination with (\ref{RGlaw2}) and
Friedmann's (\ref{FL1}), we arrive at a simple differential equation
for the matter density,
\begin{equation}\label{rhomRG}
\frac{d\rmm}{dz}-\frac{3(1+\omm)\,(1-\nu)}{1+z}\,\rmm=0\,,
\end{equation}
whose trivial integration yields\,\footnote{This equation was first
studied in \cite{oldCCstuff1}, and later was also considered  in
\cite{Wang:2004cp}.}
\begin{equation}\label{mRG}
\rho_m(z) =\rmo\,(1+z)^{3(1+\omm)(1-\nu)}\,.
\end{equation}
Here $\rmo$ is the matter density at the present time ($z=0$). The
vacuum energy density then follows from integrating once more
(\ref{Bronstein2}) using the result (\ref{mRG}):
\begin{equation}\label{CRG}
\rL(z)=\rLo+\frac{\nu\,\rmo}{1-\nu}\,\left[(1+z)^{3(1+\omm)(1-\nu)}-1\right]\,.
\end{equation}
Substituting (\ref{frie456}) in the last two equations, with
$z=(1-a)/a$, one may obtain the explicit time evolution of the
matter and vacuum energy densities, if desired. Recall that $\omm=0$
($\omm=1/3$) for non-relativistic (relativistic) matter. From
equation (\ref{mRG}) we see that, for the matter epoch, the density
does no longer evolve in the standard form $\rho_m(z)=\rmo
(1+z)^{3}$ but as $\rho_m(z)=\rmo (1+z)^{3(1-\nu)}$. The
$\nu$-correction in the power is caused by the exchange of energy
between matter and the vacuum. This is also reflected in the
corresponding non-constant behavior of $\rL(z)$ in (\ref{CRG}).
Similarly, the energy density of radiation takes on the corrected
form $\rho_R(z)=\rho_R^0 (1+z)^{4(1-\nu)}$, instead of
$\rho_R(z)=\rho_R^0 (1+z)^{4}$. Such correction to the radiation
density can be used to put a moderate limit $|\nu|<0.1$ to the basic
parameter $\nu$ in this model from primordial nucleosynthesis
\,\cite{oldCCstuff1}, although that limit is actually superseded by
better bounds derived from other precision observables which will be
discussed in the next section. Obviously, for $\nu=0$ we recover
both the standard expressions for the evolution of the matter
density in the $\CC$CDM cosmology, and the constancy of the vacuum
energy: $\rL=\rLo$. As a consistency check, we note that if we
substitute (\ref{CRG}) into (\ref{RGlaw2}), we recover the
normalized Hubble flow (\ref{nomalHflow}) for this model in the
matter dominated epoch.

The following observations are in order. Owing to the coupling
between the time-dependent vacuum and the matter component in the
$\Lambda_{t}$CDM model, there is either a particle production
process or an increase in the mass of the dark matter particles
\cite{BPS09a,AL05}. The microscopic details of this coupling are not
provided by studying the aforementioned global cosmological
properties of the model. Therefore, one has to preclude the
possibility that there can be a significant decay of vacuum energy
into matter, especially into ordinary cold and relativistic matter
(e.g. baryons and photons). This is usually \textit{not} the case
because on theoretical grounds the parameter $\nu$ in (\ref{nu}) is
naturally predicted to be small ($|\nu|\lesssim 10^{-2}$) (as it
plays the role of a small $\beta$-function), and moreover in
practice the fits to the cosmological data also confirm this fact
and enforce it to be even smaller.

If we compare the situation with the $\Lambda_{t}G_{t}$CDM running
model of the previous section, we can see that both models share the
running vacuum law (\ref{RGlaw2}) -- and hence the dependence on the
$\nu$-parameter -- but in the $\Lambda_{t}G_{t}$CDM  case  matter is
covariantly conserved, and as a result there is no danger that the
vacuum decays in excess into ordinary matter. In contrast, the
feedback associated to the time evolution of the vacuum in the
$\Lambda_{t}G_{t}$CDM model produces in this case a slow
(logarithmic) evolution of the gravitational constant with the
universe's expansion, see (\ref{GH}). We shall analyze the best fit
values of $\nu$ for the two running models $\Lambda_{t}$CDM and
$\Lambda_{t}G_{t}$CDM when we compare them with the current
precision observations in the next sections.

Finally, as the data on structure formation will be involved in
restricting the value of $\nu$, let us also quote the corresponding
differential equation for the matter density contrast
$D\equiv\delta\rho_m/\rho_m$ in the current $\Lambda_{t}$CDM model.
Denoting differentiation with respect to the cosmic time by a dot,
we have\,\cite{BPS09a}
\begin{equation}
\label{eq:11}
\ddot{D}+(2H+Q)\dot{D}-\left[4\pi\,G{\rho_{m}}-2HQ-\dot{Q}
\right]D=0\,,
\end{equation}
where the time evolving vacuum energy $\rL=\rL(t)$ affects the
growth factor via the function
\be \label{massvac} Q(t)=-\dot{\rho}_{\CC}/\rho_{m}\;. \ee
For $\rL=$const. the equation (\ref{eq:11}) reduces to the standard
form of the matter perturbations in the $\CC$CDM model, i.e.
$\ddot{D}+2H\dot{D}-4\pi\,G\rho_{m}=0$\,\cite{Peeb93}. Recall that
this equation is formally valid both for $\rL=0$ and for $\rL\neq 0$
provided $H$ in it is changed accordingly.  The numerical solution
of the more general equation (\ref{eq:11}) for the case of the
$\Lambda_{t}$CDM model has been studied in \,\cite{BPS09a}, and the
corresponding results will be used in the section
\ref{sec:ConfrontingRunningFLRW} in order to confront the running
cosmological models with the observations. Let us point out that,
for the $\Lambda_{t}$CDM model, the perturbations in $\rL$ are
subject to some ambiguities explained in Ref.\cite{FShSol07}, and
therefore for this model we shall present the matter perturbations
in the effective framework defined in equation (4.10) , in which it
is perfectly consistent to have matter perturbations for any
non-perturbed vacuum function $\rL=\rL(t)$. This is in contrast to
the $\Lambda_{t}G_t$CDM model, where perturbations in all variables
are necessary to get a consistent picture, as explained in the
previous section.

\section{Equation of state analysis of the running vacuum models}\label{sec:EOS}

A very important part in the study of the Hubble expansion behavior
of a given cosmological model is the analysis of the equation of
state (EoS) of the dark energy component. In contrast to cold and
relativistic matter, the effective EoS of the DE may appear in
general as a non-trivial function of time $\omega_{\rm eff}(t)$, or
of the cosmological redshift, $\omega_{\rm eff}(z)$. This feature
may also be expected for general CC models of the form
(\ref{varibleCCG}), for which the EoS is
$p_{\CC}(z)=-\rho_{\CC}(z)$. One can easily understand the reason as
follows -- see \cite{SS12} for details. The usual way to
parametrize the DE is to consider such entity as a self-conserved
``fluid'' at fixed Newton's coupling $G$. This representation may be
called the ``DE picture'' of the original CC model\,\cite{SS12}.
Since the CC models under consideration do \textit{not} satisfy at
least one of these two conditions (either because the vacuum energy
interacts with matter or because it evolves with time along with
$G$), the effective EoS of the system in the DE picture will be
time/redshift dependent.

Let the self-conserved fluid of the DE picture be characterized by
density and pressure $(\rs,\rD)$, therefore satisfying
$\dot{\rho}_{m s}+\alpha\,H_{\rm D}\,\rs=0$ and
$\dot{\rho}_D+3\,H_{\rm D}\,(1+\weff)\rD=0$, where
$\weff(z)=p_D(z)/\rD(z)$ is the non-trivial EoS function of the
redshift that we are looking for. The subscript D in these variables
serves us to distinguish the DE picture from the original CC picture
-that is, the original description of the model in terms of variable
cosmological parameters, which we will denote by $H_\CC$. Of course
$\rs\neq\rM$ and $\rD\neq\rL$ in general. The Hubble rate in the DE
picture takes on the form:
\begin{equation}\label{DEpicture}
H_{\rm D}^2=\frac{8\pi
G_0}{3}(\rs(z)+\rD(z))=H^2_0\,\left[\tOM^0\,(1+z)^{\alpha}
+\tOD(z)\right]\,,
\end{equation}
where
\begin{equation}\label{rhoD}
\tOD(z)=\tOD^0\,\exp\left\{3\,\int_0^z\,dz'
\frac{1+\weff(z')}{1+z'}\right\}\,,
\end{equation}
and we have defined  $\tOM^0=\rs^0/\rc^0$. Notice that the parameter
$\tOM^0$ need not coincide in general with $\OM^0$, as they are
determined from two different parameterizations of the data.
Similarly, we have introduced $\tOD(z)=\rD(z)/\rc^0$ -- and its
current value $\tOD^0=\tOD(0)$. Notice also that $G_0=1/M_P^2$ is
strictly constant, whereas $G$ in (\ref{Eg}) is in general a
variable function of the redshift: $G=G(z)$. The effective EoS in
the DE picture now follows from
\begin{equation}\label{we2}
\weff(z)=-1+\frac13\,\frac{1+z}{\rD}\,\frac{d\rD}{dz}\,,
\end{equation}
where it is to be understood that $\rD$ must be computed from the
matching condition between the DE picture and the original CC one,
i.e. by requiring that the expansion histories of the universe are
numerically equal in both pictures: $H_{\rm D}(z)=H_{\CC}(z)\,.$
Using the above formulas, let us compute the effective EoS in each
case for the matter dominated epoch.  For the running FLRW model
($\Lambda_{t}G_{t}$CDM) of section \ref{sec:runningFLRW}, we find
\begin{equation}\label{weffCGCDM}
\hspace{-2cm} \weff^{\CC G}(z)
=-1+\frac{\left(g(z)\OMo-\tOMo\right)\,(1+z)^3}{g(z)\left(\OMo
(1+z)^3+\OL(z)-\tOMo(1+z)^3\right)}\,,
\end{equation}
where $\OL(z)$ is given by (\ref{OLz}) and $g(z)$ by
(\ref{implicitgz}), and use has been made of the Bianchi identity
(\ref{bianchi2}). Similarly, for the running $\Lambda_{t}$CDM model
of section \ref{sec:runningLCDM}, we obtain
\begin{equation}\label{wpflat1}
\hspace{-2cm} \weff^{\CC}(z)
=-1+(1-\nu)\,\frac{\OMo\,(1+z)^{3(1-\nu)}-\tOMo\,(1+z)^3}
{\OMo\,[(1+z)^{3(1-\nu)}-1]-(1-\nu)\,[\tOMo\,(1+z)^3-1]}\,.
\end{equation}
In these formulas, we assume that the cosmological mass parameters
in the two pictures are in general different, i.e. $\Delta\OMo\equiv
\OMo-\tOMo\neq 0$.

\FIGURE[t]{ \centering \mbox{\epsfxsize=11cm
\epsffile{figEOS-2.eps}} \caption{The behavior of the effective EoS
function $\omega_{\rm eff}(z)$ for the two running vacuum models. In
each case, $\OMo$ corresponds to the central value obtained in our
fit (cf. Sect.~\ref{sec:ConfrontingRunningFLRW}) whereas the value
of $\nu$ is in the lower limit of the 1$\sigma$ range. For fixed
$\nu=-0.004$, and different values and signs of $\Delta\OM$, the
effective EoS behavior of both running models appears as
significantly different from the $\CC$CDM prediction
($\omega_\CC=-1$). Indeed, the effective EoS of the running $\CC$
models is seen to present quintessence- or phantom-like behavior,
depending on the values of the parameters. }\label{fig:EoS}}

It is remarkable to note that in all time-dependent vacuum models of
the form (\ref{varibleCCG}), a value $z=z^*$ {\em always} exists for
which the effective EoS $\weff(z)$ crosses the phantom divide
$\weff(z^*)=-1$, see \cite{SS12} for the general proof of this
statement. For the models under consideration we can easily check
that it is so. This is particularly simple if we look at equation
(\ref{weffCGCDM}). For, as $\OMo$ and $\tOMo$ must be very close,
and $g(z)$ is a slowly monotonous function, the effective EoS
crosses the phantom divide at some point $z^*$ characterized by the
condition $g(z^*)\OMo=\tOMo$. Whether the crossing point $z^*$ is in
the past or in the future, it will also depend on the sign of
$\Delta\OMo$ and the value of $\nu$. Clearly, for $\Delta\OMo=0$ the
crossing is exactly at $z^*=0$ since $g(0)=1$. Similarly, it is easy
to see that, for $z\to 0$, equation (\ref{wpflat1}) renders
$\weff^{\CC}\to -1$ for $\Delta\OMo=0$, whereas if $\Delta\OMo>0$ or
$\Delta\OMo<0$ the behavior is $\weff^{\CC}\gtrsim-1$
(quintessence-like) or $\weff^{\CC}\lesssim-1$ (phantom-like)
respectively. Furthermore, for both models one can easily show that,
if $\Delta\OMo=0$, the quintessence/phantom-like behavior at points
$z$ near our time is completely controlled by the sign of $\nu$.
Indeed, expanding the above expressions in the aforementioned
conditions and for small $\nu$, we meet:
\begin{eqnarray}\label{expansionsnearnow}
\hspace{-2cm} \weff^{\CC G}(z)&
=&-1-3\,\nu\frac{\OMo}{\OLo}\,(1+z)^3\,\ln\left(\OMo(1+z)^3+\OLo\right)+{\cal O}(\nu^2)\,,\\
\hspace{-2cm} \weff^{\CC}(z) &
=&-1-3\,\nu\frac{\OMo}{\OLo}\,(1+z)^3\,\ln(1+z)+{\cal O}(\nu^2)\,.
\end{eqnarray}
These two equations are very similar, but not quite. Nevertheless we
find once more that both running models share the property
$\weff=-1$ at $z=0$. Furthermore, for both running models we also
learn that, if $\Delta\OMo=0$, the cases $\nu<0$ or $\nu>0$ amount
to quintessence-like or phantom-like behaviors respectively. For
fixed $\nu$, the quintessence-like or phantom-like behavior can also
be modulated by the sign of $\Delta\OM$. Some numerical examples are
considered in Fig.\,\ref{fig:EoS}, where the crossing is in the
future, except for the cases $\Delta\OMo=0$ in which the crossing is
at $z=0$ (i.e. precisely at our time).

The previous discussion clearly illustrates, with the help of the
two non-trivial running models $\Lambda_{t}G_{t}$CDM and
$\Lambda_{t}$CDM discussed in the previous sections, that a generic
cosmological model (\ref{varibleCCG}) with time-varying vacuum
energy generally leads, when described in the DE picture (i.e. as if
it consisted of a cosmological self-conserved DE fluid at constant
$G$), to a crossing of the phantom divide. Thus it may provide a
natural explanation for the observational data, which still admit a
tilt in the phantom domain\,\cite{komatsu08}, or it may simply
explain the apparent quintessence-like behavior, despite there is no
fundamental quintessence or phantom field in our framework. In fact,
remember that the original CC picture for both models is just a
time-varying vacuum model with or without interaction with matter.

In the next section, we confront these running models with the
observational data on SNIa, CMB and BAO. As we will see, for values
of $\nu$ sufficiently large within the 1$\sigma$ range of our fits,
both models predict a cluster redshift distribution that differs
noticeably from the $\CC$CDM expectations. For such values of $\nu$,
the EoS function of our models should also be easily distinguishable
from the $\CC$CDM value, $\omega_\CC=-1$, even in the simplest
$\Delta\OMo=0$ case. We refer once more to Fig.\,\ref{fig:EoS},
where we have plotted the EoS functions (\ref{weffCGCDM}) and
(\ref{wpflat1}). For large redshifts ($z\sim1.5-2$) it is even
conceivable that the quality of the data provided by forthcoming
SNIa experiments could be enough so as to discriminate between both
running cosmological models. A combined analysis of clusters and the
dark energy EoS could be a most effective strategy to put our models
to the test.

\section{Confronting the running vacuum models with the latest
observations}\label{sec:ConfrontingRunningFLRW} In this section, we
briefly present the basic observational samples and data statistical
analysis that will be adopted to constrain the ``running'' models
presented in the previous sections. First of all, we use the {\em
Constitution} set of 397 type Ia supernovae of Hicken et al.
\cite{Hic09}. In order to avoid possible problems related with the
local bulk flow, we use a subsample of 366 SNIa, excluding those
with $z<0.02$. The corresponding $\chi^{2}_{\rm SNIa}$ function, to
be minimized, is:
\begin{equation}
\label{chi22} \chi^{2}_{\rm SNIa}({\bf p})=\sum_{i=1}^{366} \left[
\frac{ {\cal \mu}^{\rm th} (a_{i},{\bf p})-{\cal \mu}^{\rm
obs}(a_{i}) } {\sigma_{i}} \right]^{2} \;,
\end{equation}
where $a_{i}=(1+z_{i})^{-1}$ is the observed scale factor of the
Universe for each data, $z_{i}$ is the observed redshift, ${\cal
\mu}$ is the distance modulus ${\cal \mu}=m-M=5{\rm log}d_{L}+25$
and $d_{L}(a,{\bf p})$ is the luminosity distance:
\begin{equation}
d_{L}(a,{\bf p})=\frac{c}{a} \int_{a}^{1} \frac{{\rm
d}a'}{a'^{2}H(a')} \;,
\end{equation}
with $c$ the speed of light and ${\bf p}$ a vector containing the
cosmological parameters (i.e., $\Omega_{m}^{0},\nu$) that we wish to
fit for. The previous formula applies only for spatially flat
universes, which we are assuming throughout.

On the other hand, we consider the baryonic acoustic
oscillations (BAOs), which are
produced by pressure
(acoustic) waves in the photon-baryon plasma of the early universe,
as a result of the existence of dark matter overdensities. Evidence of this excess has
been found in the clustering properties of the SDSS galaxies
(see \cite{Eis05}, \cite{Perc10}) and it provides a ``standard ruler'' we can
employ to constrain dark energy models. In this work we use
the newest results of Percival et al. \cite{Perc10},
$r_{s}(z_{d})/D_{\rm V}(z_{\star})=0.1390\pm 0.0037$ (see also
\cite{Kazin10}). Note that
$r_{s}$ is the comoving sound horizon
size at the baryon drag epoch\footnote{$z_{d}$ is given by the
fitting formula of \cite{Eis98}.} $z_{d}$,
$D_{V}(z)$ is the effective distance measure
\cite{Eis05} and $z_{\star}=0.275$. Of course, the
quantities $(r_{s},D_{\rm V})$ can be defined analytically.
In particular, $r_{s}$ is given by:
\begin{equation}
r_{s}(z_{d})=\frac{c}{\sqrt{3}}\int_{0}^{a_{d}}
\frac{da}{a^{2} H(a) \sqrt{1+(3\Omega_{b}^{0}/4\Omega_{\gamma}^{0})a } }\;,
\end{equation}
where $a_{d}=(1+z_{d})^{-1}$ and $\Omega_{b}^{0}h^{2}\simeq 0.02263$.
In this context, the effective distance is (see \cite{Eis05}):
\begin{equation}
D_{\rm V}(z)\equiv \left[ (1+z)^{2} D_{A}^{2}(z) \frac{cz}{H(z)}\right]^{1/3}\;,
\end{equation}
where $D_{A}(z)=(1+z)^{-2} d_{L}(z,{\bf p}) $ is the angular diameter
distance. Therefore, the corresponding
$\chi^{2}_{\rm BAO}$ function is simply written as:
\begin{equation}
\chi^{2}_{\rm BAO}({\bf p})=
\frac{\left[\frac{r_{s}(z_{d})}{D_{\rm V}(z_{\star})}({\bf
p})-0.1390\right]^{2}}{0.0037^{2}} \;.
\end{equation}
\begin{figure}[t]
\begin{center}
\mbox{\epsfxsize=14.5cm \epsffile{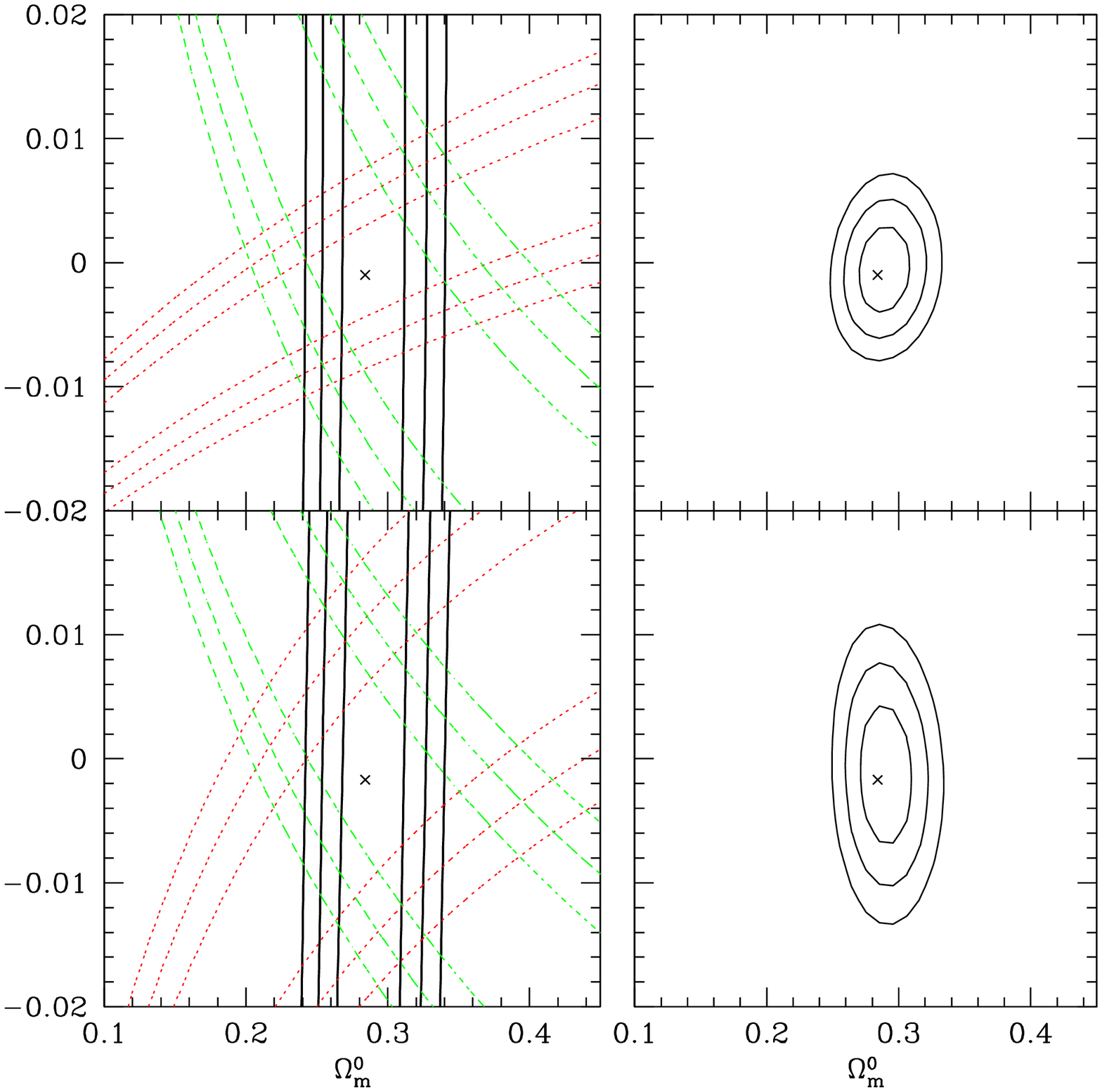}} \caption{{\it Upper
Panel:} Likelihood contours (for $-2{\rm ln}{\cal L}/{\cal L}_{\rm
max}$ equal to 2.30, 6.16 and 11.81, corresponding to 1$\sigma$,
2$\sigma$ and $3\sigma$ confidence levels) in the
$(\Omega_{m}^{0},\nu)$ plane for the $\CC_t G_t$CDM model. The left
panel shows the contours based on the SNIa data (thick solid black
lines), BAOs (dotted-red lines) and CMB shift parameter
(dashed-green lines). In the right panel we show the corresponding
contours based on the joint statistical analysis (SNIa+BAO+CMB
data). {\it Lower Panel:} Similar likelihood contours for the
$\Lambda_{t}$CDM model.}\label{fig:fig1}
\end{center}
\end{figure}

Finally, a very accurate and deep geometrical probe of
dark energy is the angular scale of the sound horizon at the last
scattering surface, as encoded in the location $l_1^{TT}$ of the
first peak of the Cosmic Microwave Background (CMB) temperature
perturbation spectrum. This probe is described by the  CMB shift
parameter \cite{Bond:1997wr,Nesseris:2006er}, defined as:
\begin{equation}
R=\sqrt{\Omega_{m}}\int_{a_{ls}}^1 \frac{da}{a^2 E(a)}
\end{equation}
The measured shift parameter according to the WMAP 7-years data \cite{komatsu08} is
$R=1.726\pm 0.018$ at $z_{ls}=1091.36$
[or
$a_{ls}=(1+z_{ls})^{-1}\simeq 9.154\times 10^{-4}$].
In this case, the $\chi^{2}_{\rm cmb}$
function is given by:
\begin{equation}
\chi^{2}_{\rm cmb}({\bf p})=\frac{[R({\bf p})-1.726]^{2}}{0.018^{2}}\;.
\end{equation}
Note that the measured CMB shift parameter is somewhat model
dependent, but mainly for models that include massive neutrinos or
those with a strongly varying equation of state parameter (which is
not our case). For a detailed discussion of the shift parameter as a
cosmological probe, see \cite{Elgaroy07}.

In order to place tighter constraints on the corresponding
parameter space of our model, the probes described above must be combined
through a joint likelihood analysis\footnote{Likelihoods are
normalized to their maximum values. In the present analysis we
always report $1\sigma$ uncertainties on the fitted parameters. Note
also that the total number of data points used here is
$N_{tot}=368$, while the associated degrees of freedom is: {\em
  dof}$= N_{tot}-n_{\rm fit}$, where $n_{\rm fit}$ is
the model-dependent number of fitted
parameters.}, given by the product of the individual likelihoods
according to:
\be
{\cal L}_{tot}({\bf p})= {\cal L}_{\rm SNIa}\times
{\cal L}_{\rm BAO} \times {\cal L}_{\rm cmb}\;,
\ee
which translates in an addition for the joint $\chi^2$ function:
\be
\chi^{2}_{tot}({\bf p})=\chi^{2}_{\rm SNIa}+\chi^{2}_{\rm
BAO}+\chi^{2}_{\rm cmb}\;.
\ee
In our $\chi^2$ minimization procedure,
we use the following range and steps of the fitted parameters:
$\Omega_{m}^{0} \in [0.1,1]$ in steps of 0.001
and $\nu \in [-0.02,0.02]$  in steps of
$10^{-4}$.

In Figure \ref{fig:fig1} we present the results of our analysis for
the $\Lambda_{t}G_{t}$CDM (upper panel) and the $\Lambda_{t}$CDM
(lower panel) models. The left plots in that figure show the
1$\sigma$, 2$\sigma$ and $3\sigma$ confidence levels in the
$(\Omega_{m}^{0},\nu)$ plane, with the SNIa-based results indicated
by thick solid lines, the BAO results by dotted - red lines and
those based on the CMB shift parameter by dashed -green lines. Using
the SNIa data alone it is evident that although the $\Omega_{m}^{0}$
parameter is tightly constrained ($\simeq 0.284$), the $\nu$
parameter remains completely unconstrained. As can be seen in the
right plots of Figure \ref{fig:fig1}, the above degeneracy is broken
when using the joint likelihood analysis, involving all the
cosmological data. For the $\Lambda_{t}G_{t}$CDM we find that the
overall likelihood function peaks at
$\Omega_{m}^{0}=0.283^{+0.012}_{-0.013}$ and $\nu=-0.001\pm 0.0032$
with $\chi_{tot}^{2}(\Omega_{m}^{0},\nu) \simeq 439.36$ for $366$
degrees of freedom.

On the other hand, for the $\Lambda_{t}$CDM model the best fit
parameters are: $\Omega_{m}^{0}=0.284^{+0.011}_{-0.014}$ and
$\nu=-0.0017^{+0.0051}_{-0.0043}$ with
$\chi_{tot}^{2}(\Omega_{m}^{0},\nu)/dof \simeq 439.34/366$. Finally,
for the usual $\Lambda$ cosmology ($\nu=0$ and $G=G_{0}$) we find
that the overall likelihood function peaks at
$\Omega_{m}^{0}=0.284\pm 0.012$ with
$\chi_{tot}^{2}(\Omega_{m}^{0})/dof\simeq 439.61/367$.
%Based on the previous results we adopt, throughout the remaining
%paper, $\Omega_{m}^{0}=0.28$.
We remark that although for both running models the best fit value
for the $\nu$ parameter is negative\,\footnote{Notice that in
Ref.\cite{BAS09} the best fit value for the $\Lambda_{t}$CDM model
was $\nu\simeq 0.002>0$, because $\nu$ was actually sampled only
within the interval $\nu\in[0,0.3]$. The value of $\OMo$, however,
remains very similar.}, its $1\sigma$ range includes both positive
and negative $\nu$ values and encompasses quite a wide segment. For
instance, for the $\Lambda_{t}G_{t}$CDM model the 1$\sigma$ range is
roughly $[-0.004, 0.002]$. It is worth mentioning that this range of
values is in the same ballpark as the bounds obtained for $\nu$ from
the normalization of the spectrum amplitude and from primordial
nucleosynthesis\,\cite{GSFS10}.

To end this section, let us mention that we have checked that using
the earlier BAO results of Eisenstein et al. \cite{Eis05} does not
change significantly the previously presented constraints.

%\section{Application to the formation of
%collapsed structures}\label{sec:RunningFLRWCollapse}

\section{Halo abundances and their evolution in dark energy
models}\label{sec:RunningFLRWCollapse} In an attempt to define
observational criteria that will enable us to identify the realistic
variants of the concordance $\Lambda$CDM cosmology and distinguish
among the different $\Lambda(t)$ models, we derive and compare their
theoretically predicted cluster-size halo redshift distributions.

The formalism to compute the fraction of matter in the universe that
has formed bounded structures and its redshift distribution was
developed in 1974 by Press and Schechter \cite{press} (hereafter
PSc). In their approach, the primordial density fluctuation for a
given mass $M$ of the dark matter fluid is described by a random
Gaussian field.  One introduces a function, ${\cal F}(M,z)$,
representing the fraction of the universe that has collapsed by the
redshift $z$ in halos above some mass $M$. With this  function  one
may estimate the (comoving) number density of halos, $n(M,z)$, with
masses within the range $(M, M+\delta M)$: \be n(M,z) dM=
\frac{\partial {\cal F}(M,z)}{\partial M} \frac{{\bar \rho}}{M} dM.
\ee This expression can be rewritten as follows:
\begin{eqnarray}\label{MF}
n(M,z) dM &=& -\frac{\bar{\rho}}{M} \left(\frac{1}{\sigma} \frac{d
\sigma}{d M}\right) f_{\rm PSc}(\sigma) dM  = \frac{\bar{\rho}}{M}
\frac{d{\rm \ln}\sigma^{-1}}{dM} f_{\rm PSc}(\sigma) dM,
\end{eqnarray}
where $f_{\rm PSc}(\sigma)=\sqrt{2/\pi} (\delta_c/\sigma)
\exp(-\delta_c^2/2\sigma^2)$, $\delta_{c}$ is the linearly
extrapolated density threshold above which structures collapse
\cite{eke} (see Appendix ~\ref{sec:app} for more details), while
$\sigma^2(M,z)$ is the mass variance of the smoothed linear density
field, which depends on the redshift $z$ at which the halos are
identified. It is given in Fourier space by: \be \label{sig88}
\sigma^2(M,z)=\frac{D^2(z)}{2\pi^2} \int_0^\infty k^2 P(k) W^2(kR)
dk \;, \ee where $D(z)$ is the growth factor of
perturbations\footnote{The equations to compute the growth factor
have been given at the end of Sects.\,\ref{sec:runningFLRW} and
\ref{sec:runningLCDM}, namely Eqs.\,(\ref{supeq}) and (\ref{eq:11}).
Since the pure matter universe (Einstein de-Sitter) has the solution
$D_{\rm EdS}=a$, we normalize our vacuum models such as to get
$D\simeq a$ at large redshifts (e.g. $z=500$), where the matter
component dominates the cosmic fluid and our models resemble a pure
CDM model.}, $P(k)$ is the power-spectrum of the linear density
field, and $W(kR)=3({\rm
  sin}kR-kR{\rm cos}kR)/(kR)^{3}$ is the top-hat smoothing function,
which contains on average a mass $M$ within a radius $R=(3M/ 4\pi
\bar{\rho})^{1/3}$ and $\bar{\rho}=2.78 \times
10^{11}\Omega_{m}^{0}h^{2}M_{\odot}$Mpc$^{-3}$. We use the CDM power
spectrum:
\begin{equation}
P(k)=P_{0} k^{n} T^{2}(\Omega_{m}^{0},k)\,,
\end{equation}
with $T(\Omega_{m}^{0},k)$ the BBKS transfer function \cite{Bard86}:
%%%%%%%%%%%%%%%%%%%%%%%
\begin{eqnarray} \label{jtf}
T(\Omega_{m}^{0},k)&=&\frac{\ln (1+2.34 q)}{2.34 q}\Big[1+3.89 q + (16.1 q)^2 + \, (5.46 q)^3+(6.71 q)^4\Big]^{-1/4}\,,\nonumber\\
q=q(k)&\equiv&\frac{k}{h \Gamma}
%=frac{k}{\OMo\,h^2\,  e^{-\,\Omega_b^0\,-\sqrt{2h}\,\left({\Omega_b^0}/{\OMo}\right)}}
\,.
\end{eqnarray}
%%%%%%%%%%%%%%%%%%%%%%%
Here $\Gamma$ is the shape parameter, according to \cite{Sugi},
provided by the 7-year WMAP results \cite{komatsu08}, in combination
with the model fitted $\Omega_m$ values of section 6.
%{and parameters ($n$, $\Omega_b$, $h$) provided by the 7-year WMAP results \cite{komatsu08}.}
Note that in this approach all the mass
is locked inside halos, according to the normalization constraint:
\be \int_{-\infty}^{+\infty} f_{\rm PSc}(\sigma) d{\rm
\ln}\sigma^{-1} = 1\;. \ee
%We remind the reader that it is
It is traditional to parametrize the mass variance in terms of
$\sigma_8$, the rms mass fluctuation amplitude on scales of $R_{8}=8
\; h^{-1}$ Mpc at redshift $z=0$ [$\sigma_{8} \equiv \sigma_8(0)$],
so that we have: \be \label{s888} \sigma^2(M,z)=\sigma^2_8(z)
\frac{\int_{0}^{\infty} k^{n+2} T^{2}(\Omega_{m}^{0}, k) W^2(kR) dk}
{\int_{0}^{\infty} k^{n+2} T^{2}(\Omega_{m}^{0}, k) W^2(kR_{8})
dk}\,, \ee where \be \sigma_8(z)=\sigma_8\frac{D(z)}{D(0)} \;. \ee

Although the previously described Press-Schechter formalism was
shown to provide a good first approximation to the halo mass
function obtained by numerical simulations, it was later found to
over-predict/under-predict the number of low/high mass halos at the
present epoch \cite{Jenk01,LM07}. More recently,  a large number of
works have provided better fitting functions for $f(\sigma)$, some
of them  based on a phenomenological approach. In the present
treatment, we adopt the one proposed by  Reed et al. \cite{Reed}:
\begin{equation}\label{neffeqq}
f(\sigma, n_{\rm eff}) = A\sqrt{{2b\over\pi}}
\left[1+\left({\sigma^2\over b\delta_c^2}\right)^p + 0.6G_1 + 0.4G_2\right]{\delta_c\over\sigma}
\exp\left[-{cb\delta_c^2\over2\sigma^2}
-{0.03 \over (n_{\rm eff}+3)^2} \left({\delta_c \over \sigma}
\right)^{0.6} \right],\nonumber\\
\end{equation}
where $A = 0.3222,\, p = 0.3,\, b = 0.707,\, c=1.08$, while
$G_1,\,G_2$ and $n_{\rm eff}$, {the slope of the non-linear power-spectrum at
the halo scale}, are given by:
\begin{equation}
G_1 = \exp\left[-{{(\ln\sigma^{-1}-0.4)^2} \over {2(0.6)^2}}\right],\;
G_2 = \exp\left[-{{(\ln\sigma^{-1}-0.75)^2} \over
    {2(0.2)^2}}\right],\;
n_{\rm eff} = 6 {{\rm d}\ln\sigma^{-1}\over{\rm d}\ln M\phantom{+}} -3.
\end{equation}
%{\bf with $\delta_{c} \simeq 1.675$ for the $\Lambda$ cosmology
%\cite{Wein03}.}
%and the WMAP7 normalization of the power-spectrum ($\sigma_8\simeq 0.8$).

%\begin{figure}[ht]
%\mbox{\epsfxsize=8.5cm \epsffile{FIG1.ps}} \caption{The halo mass
%function at two different redshifts. The different DE models are
%represented by different symbols and/or line types (see Table 1
%for definitions).}
%\end{figure}

\subsection{Collapse threshold and mass variance of the running vacuum models}\label{sec:RunningFLRWCollapse1}
In order to compare the mass function predictions for the different
vacuum models, it is imperative to use for each model the
appropriate value of $\delta_c$ and $\sigma_8$. Indeed, the way in
which dark energy (time-varying vacuum energy, in our case) affects
the formation of gravitationally bound systems (clusters of
galaxies) is a crucial question that has received a lot of attention
in the literature. Three distinct scenarios (with increasing level
of difficulty in the theoretical treatment) can be considered, to
wit: (i) dark energy remains homogeneous and only matter virializes;
(ii) dark energy may cluster but only matter virializes; and (iii)
dark energy may also cluster and the whole system (matter and dark
energy) virializes. The first scenario has been widely used in the
literature, probably because it is the simplest one\,
\cite{Mota04,Nunes06,Basi07,Pace10,Schmidt10}. In the present paper,
we will consider just this canonical scenario when dealing with the
$\Lambda_t$CDM model. For the full running FLRW model (or $\Lambda_t
G_t$CDM model), however, which is the main object of the present
study, we will go a bit deeper and we shall treat the gravitational
collapse within the framework of the second scenario (in which
$\delta G$ and $\delta_\Lambda$ are both non-vanishing and only
matter virializes). Then, in the Appendix ~\ref{sec:app} we shall
compare these results for the $\Lambda_t G_t$CDM model within
scenario (ii) to those obtained within the canonical scenario (i),
for which $\delta G=\delta_\Lambda=0$\, \footnote{A complete study
of all the possible clustering scenarios in the context of our
vacuum models is beyond the scope of the present paper. Scenario
(iii) is of course the most difficult one, and although a first
attempt to include the effects of clustered and globally virialized
vacuum energy and matter for the $\Lambda_t$CDM model has been
considered in \cite{BPS10a}, the modeling of this situation requires
the introduction of new assumptions and more free parameters. In
order to avoid a too cumbersome treatment here, the corresponding
study for the $\Lambda_t G_t$CDM model will be presented
elsewhere.}.

It is well known that for the usual $\Lambda$ cosmology $\delta_{c}
\simeq 1.675$, while Weinberg \& Kamionkowski \cite{Wein03} provide
an accurate fitting formula to estimate $\delta_{c}$ for DE models
with constant EoS parameter. It should furthermore be noted that
such a formula is, in principle, only applicable to ``conventional''
DE models, i.e. models with self-conserved matter and DE densities
and constant $G$. Although these conditions are not satisfied by our
running vacuum models, one can show\,\cite{Pace10,BPL10} that the
$\delta_c$ values for a large family of dark energy models with a
time-varying EoS parameter can be well approximated using the
previously discussed fitting formula, as long as the EoS parameter
is not very different from -1 near the present epoch. This is indeed
the case for our models, as shown in Fig.\,\ref{fig:EoS}.
%, which is just the case for our
%running models\,\footnote{Indeed, recall from Sect. \ref{sec:EOS}
%that we can always perform a change of \emph{picture} and describe
%the running vacuum energy as if it were a self-conserved fluid,
%provided we introduce a time-dependent effective EoS $\omega_{\rm
%eff}(t)$ for the DE component in the new picture. Moreover, for the
%current vacuum models $\omega_{\rm eff}(t)$ becomes sufficiently
%close to $-1$ near our time so as to use the estimates of
%$\delta_{c}$ from\, \cite{Wein03}.}.
In spite of this prognosis, we will nevertheless carefully compute
the value of $\delta_c$ within our vacuum models, following the
prescriptions given in \cite{Pace10} (cf. Sect.~2.1 in that
reference) and \cite{Abramo}. Starting from a Newtonian formalism,
we derive a non-linear second-order differential equation for the
evolution of the matter perturbations in our vacuum models from
which we compute  $\delta_{c}(z)$, and in particular
$\delta_{c}\equiv \delta_c(0)$ (cf. Appendix ~\ref{sec:app} for the
calculational details and methodology). The resulting values are
listed in Table 1. As an example, in the case of the
$\Lambda_{t}G_{t}$CDM model, for $\Omega_{m}^{0}=0.283$ and
$\nu=-0.001$ we find $\delta_{c}\simeq 1.677$, pretty close to the
$\Lambda$CDM value ($\simeq 1.675$). On the other hand, for the
$\Lambda_{t}$CDM model ($\Omega_{m}^{0}=0.284$, $\nu=-0.0017$) we
find a bit larger value, $\delta_{c}\simeq 1.685$. In all cases,
however, the differences with respect to the $\CC$CDM value are at
the few per mil level only. This is in agreement with the
expectations for our models, and it is exactly the same that happens
in other models\,\cite{Pace10,BPL10}.

The relevant $\sigma_8$ value for the different ``running'' vacuum
models can be estimated by scaling the present time $\CC$CDM
value\footnote{In the following discussion, the quantities referred
to the $\CC$CDM model are distinguished by the subscript `$\CC$'
($\sigma_{8,\CC}$; $D_\CC$; $\Omega_{m,\Lambda}^{0}$) whereas the
corresponding quantities in the running models carry no subscript.}
($\sigma_{8, \Lambda}$) using equation (\ref{sig88}). {In general,
for any DE model, equation (\ref{sig88}) takes the form: \be
\sigma_{\rm 8}=\sigma_{8, \Lambda} \frac{D(0)}{D_{\Lambda}(0)}
\left[\frac{P_{0} \int_{0}^{\infty} k^{n+2} T^{2}(\Omega_{m}^{0},k)
W^2(kR_{8}) dk} {P_{\Lambda,0}\int_{0}^{\infty} k^{n+2}
T^{2}(\Omega_{m, \Lambda}^{0},k) W^2(kR_{8}) dk}
\right]^{1/2}\,,\label{s88general} \ee where
$P_{0}/P_{\Lambda,0}=(\Omega_{m,\Lambda}^{0}/\Omega_{m}^{0})^{2}$.}
%Since, however, for the current vacuum models we find
%$\Omega_{m}^{0}=\Omega_{m,\Lambda}^{0}$ (see section 5), the formula
%(\ref{s88general}) boils down to
%\begin{equation}\label{s88}
%\sigma_{8}=\sigma_{8, \Lambda} \frac{D(0)}{D_{\Lambda}(0)}\;.
%\end{equation}

Therefore, using the observationally determined value for
$\sigma_{8,\Lambda}$, we can easily derive the corresponding
$\sigma_{\rm 8}$ value for any of the time-varying vacuum models.
Indeed, the joint WMAP7+BAO+$H_{0}$ analysis of
Komatsu \emph{et al.} \cite{komatsu08} %(see also \cite{Dun09})
resulted in a value of $\sigma_{8, \Lambda}=0.811$. This value is in
quite good agreement with that provided by a variety of different
methods; for example, an analysis based on cluster abundances gave
the degenerate combination: $\sigma_{8, \Lambda}= \left(0.83 \pm
0.03\right) (\Omega_{m}^{0}/0.25)^{-0.41}$ \cite{Rozo09}.
%which in our case
%($\Omega_{m}^{0}=0.284$) implies $\sigma_{8, \Lambda}\simeq 0.788$.
The weak-lensing analysis of Fu et al. \cite{Fu08}
provided $\sigma_{8, \Lambda}=
\left(0.837\pm 0.084\right)(\Omega_{m}^{0}/0.25)^{-0.53}$,
%which implies {\bf in our case} $\sigma_{8, \Lambda}\simeq 0.782$,% for
                                % $\Omega_{m}^{0}=0.284$.
while two recent studies based on a joint analysis of the large-scale clustering
of red SDSS galaxies, CMB, SNIa and BAO data resulted in: $\sigma_{8,\Lambda}\simeq 0.8
\pm 0.02$ (for $\Omega^0_m\simeq 0.26$) \cite{Sanchez, Ferramacho}.
The only method that provides discrepant results is that
based on galaxy or cluster peculiar velocities \cite{Pike05, Feld03}.

Finally, inserting the Komatsu et al. \cite{komatsu08}
$\sigma_{8,\Lambda}$ value in equation (\ref{s88general}) we can
estimate the corresponding $\sigma_{\rm 8}$ values for our models.
These values will be used in the mass function analysis and are
listed in Table 1.

%It should also be mentioned that some recent papers, based on studies
%of  bulk flows on scales of $> 100 \; h^{-1}$Mpc,  suggest
%significantly higher values of $\sigma_{8}$.
%For example, the recent study of
%Watkins et al. \cite{Wat09} found a $\sigma_{8}$ normalization
%which is a factor of $\sim 2$ larger than that of the $\Lambda$CDM model.
%If these results are correct then they would strongly challenge the
%$\Lambda$CDM model. However, such a discussion is beyond the scope of
%the current study.

\subsection{Halo mass function \& number counts of the running vacuum
models}\label{sec:HaloMasFunction} 
\begin{figure}[ht]
\begin{center}
\mbox{\epsfxsize=14cm \epsffile{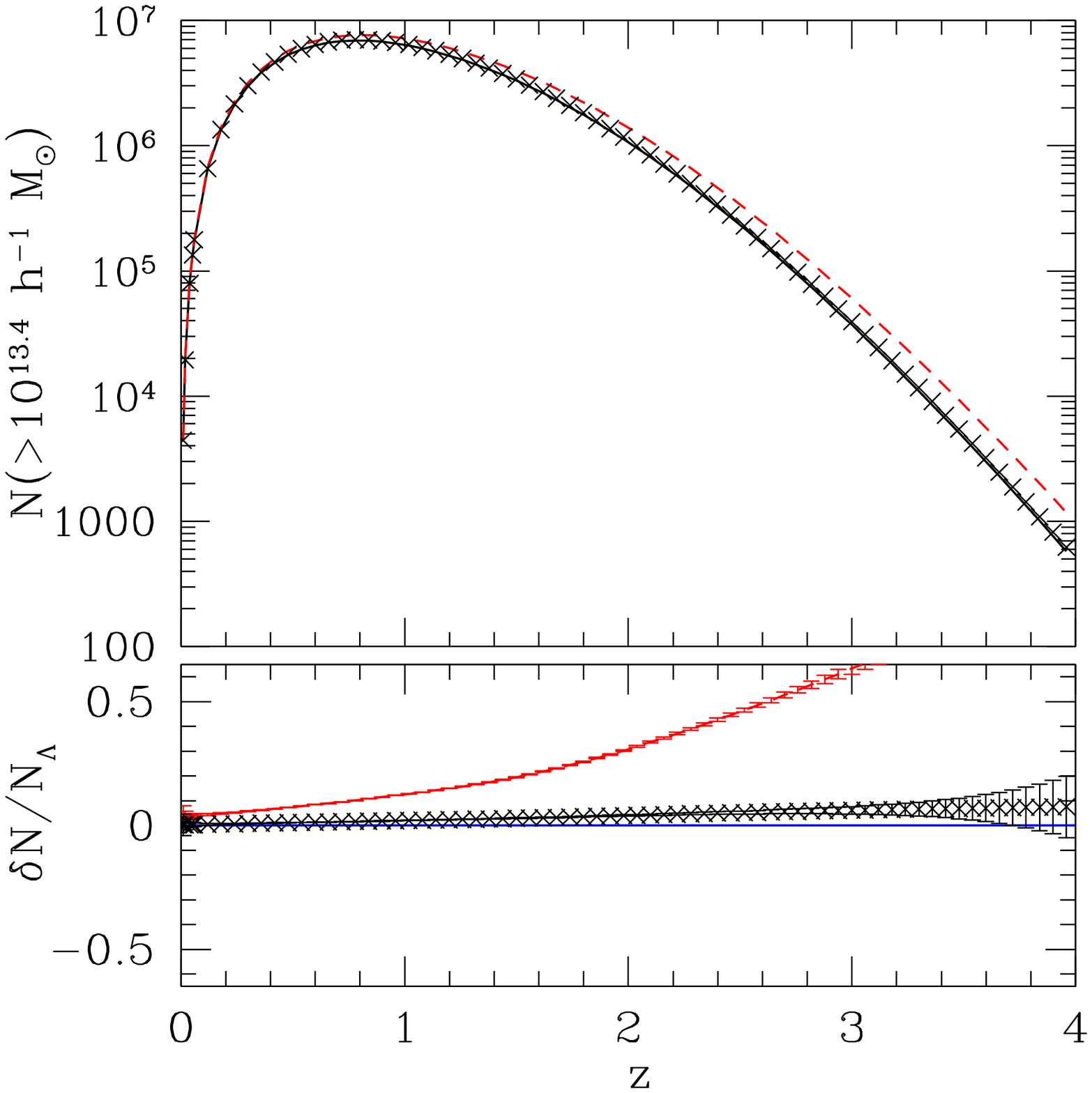}} \caption{The theoretical
redshift distribution of $M\magcir 10^{13.4} h^{-1} M_{\odot}$
clusters (upper panel) for the different vacuum models and  the
corresponding fractional difference (lower panel) between the
running models and the reference $\Lambda$CDM model.
%The lower panel shows the corresponding relative differences with respect to the
%$\Lambda$CDM model indicated in the upper panel by a black solid line.
The running models are plotted for the central fit values of the
$\nu$ parameter, $\nu=-0.0017$ (red dashed line) for the
$\Lambda_t$CDM and $\nu=-0.001$ (black crosses) for the
$\Lambda_{t}G_{t}$CDM. The full list of input parameters,
corresponding to the central fit values, as well as the symbols or
line-types characterizing the different models, are shown in Table
1.
%The values of the input parameters are indicated in detail in Table1.
Error bars are 2$\sigma$ Poisson uncertainties.}\label{figs:fig2}
\end{center}
\end{figure}

We now move to our results. Given
the halo mass function from Eq.\,(\ref{MF}) we can derive an
observable quantity which is the redshift distribution of clusters,
${\cal N}(z)$, within some determined mass range, say $M_1\le M\le
M_2$. This can be estimated by integrating the expected differential
halo mass function, $n(M,z)$, with respect to mass, according to:
\be {\cal N}(z)=\frac{dV}{dz}\;\int_{M_{1}}^{M_{2}} n(M,z)dM, \ee
where $dV/dz$ is the comoving volume element, which in a flat
universe takes the form:
\be \frac{dV}{dz} =4\pi r^{2}(z)\frac{dr(z)}{dz}, \ee with $r(z)$
denoting the comoving radial distance out to redshift $z$: \be
r(z)=\frac{c}{H_{0}} \int_{0}^{z} \frac{dz'}{E(z')}.
%\;\;\;\;\;\frac{dr}{dz}=\frac{c}{H_{0}E(z)}
\ee
In the upper panel of Fig.~\ref{figs:fig2}, we show the
theoretically expected redshift distribution, ${\cal N}(z)$, for
cluster-size halos, ie., $M_1=10^{13.4} \;h^{-1} M_{\odot}$ and
$M_2=10^{16} \;h^{-1} M_{\odot}$ for the $\Lambda_{t}G_{t}$CDM,
$\Lambda_{t}$CDM and $\CC$CDM models, using the best-fit values for
the $\Omega_m^0$ and $\nu$ parameters provided in section 6 (also
indicated in Table 1). In the lower panel we show the relative
differences of the two running models with respect to the
concordance $\Lambda$CDM model. The different models are
characterized by the symbols and line types presented in Table 1.

%It is evident that both ``running'' vacuum models show significant differences with respect to
%the concordance $\Lambda$ model, with the  $\Lambda_{t}G_{t}$CDM
%producing, as a function of redshift, larger numbers
%and the $\Lambda_{t}$CDM producing lower numbers of cluster-size halos, respectively.

It is evident that, for the central fit values of the $\nu$
parameter, the $\Lambda_{t}$CDM is the only model which shows
significant differences with respect to the concordance $\Lambda$CDM
model, producing a larger number of cluster-size halos at any given
redshift. As for the $\Lambda_{t}G_{t}$CDM model, although it also
produces larger numbers of halos than the $\Lambda$CDM model, the
differences are minimal. For instance, for $2<z<3$ the
$\Lambda_{t}$CDM model presents $\delta {\cal N}/{\cal N}\sim 0.47$,
whereas in the $\Lambda_{t}G_{t}$CDM case the corresponding value amounts
to only $\sim 0.06$.

However, in order to assess whether it is possible to
observationally discriminate between the different vacuum models, we
need to concentrate on a more significant model comparison by
deriving the expected halo redshift distributions in the context of
realistic future cluster surveys.
%%%%%%%%%%%%%%%%%%%%%%%%%%%%%%%%%%%%%%%
Two of these realistic future surveys are:

\noindent (a) The {\tt eROSITA} satellite X-ray survey, with a flux
limit of $f_{\rm lim}=3.3\times 10^{-14}$ ergs s$^{-1}$ cm$^{-2}$,
at the energy band 0.5-5 keV and covering $\sim 20000$ deg$^{2}$ of
the sky.

\noindent (b) The South Pole Telescope (SPT) Sunyaev-Zeldovich (SZ)
survey, with a limiting flux density at $\nu_0=150$ GHz of
$f_{\nu_0, {\rm lim}}=5$ mJy and a sky coverage of $\sim 4000$
deg$^{2}$.

\begin{figure}[ht]
\begin{center}
\mbox{\epsfxsize=14.5cm \epsffile{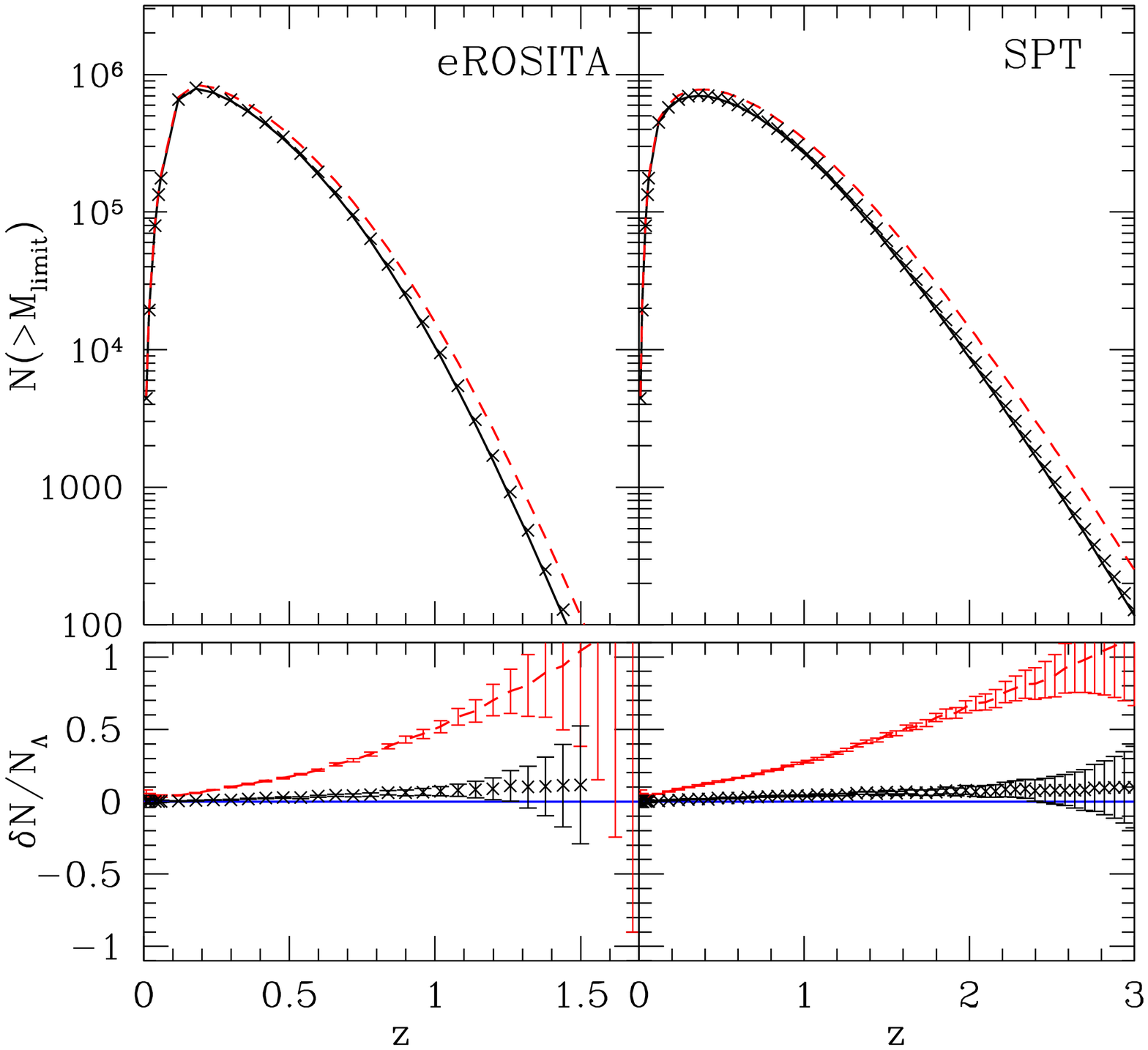}} \caption{The expected
cluster redshift distribution, {over the whole sky}, of the two
running vacuum models for the case of the two future cluster surveys
{\tt eROSITA} and SPT (upper panels), and the corresponding
fractional difference with respect to the reference $\Lambda$CDM
model (lower panels). The input parameters, corresponding to the
central fit values, as well as the symbols or line-types
characterizing the different models are shown in Table 1.}
%as in Fig.\,3, for the central fit values of
%the $\nu$ parameter, see Table 1 for more details.}
\label{figs:fig3}
\end{center}
\end{figure}

To realize the predictions of the first survey we use the relation
between halo mass and bolometric X-ray luminosity, as a function of
redshift, provided in \cite{Fedeli}: \be\label{bolom} L(M,z)=3.087
\times 10^{44} \left[\frac{M E(z)}{10^{15} h^{-1}
    M_{\odot}} \right]^{1.554} h^{-2} \; {\rm erg s^{-1}} \;.
\ee The limiting halo mass that can be observed at redshift $z$ is
then found by inserting in the above equation the limiting
luminosity, given by: $L=4 \pi d_L^2 f_{\rm lim}${\em c}$_b$, with
$d_L$ the luminosity distance corresponding to the redshift $z$ and
{\em c}$_b$ the band correction, necessary to convert the bolometric
luminosity of Eq.\,(\ref{bolom}) to the 0.5-5 keV band of {\tt
eROSITA}. We estimate this correction by assuming a Raymond-Smith
\cite{RS77} plasma model with a metallicity of 0.4$Z_{\odot}$, a typical
cluster temperature of $\sim 4$ keV and a Galactic absorption column
density of $n_{H}=10^{21}$ cm$^{-2}$.

The predictions of the second survey can be realized using again the
relation between limiting flux and halo mass from \cite{Fedeli}:
\be\label{sz} f_{\nu_0, {\rm lim}}= \frac{2.592 \times 10^{8} {\rm
mJy}}{d_{A}^{2}(z)} \left(\frac{M}{10^{15} M_{\odot}}\right)^{1.876}
E^{2/3}(z) \; \ee where $d_A(z) \equiv d_L/(1+z)^2$ is the angular
diameter distance out to redshift $z$.

In Fig.\ref{figs:fig3} (upper panels) we present the expected
redshift distributions above a limiting halo mass, which is $M_1
\equiv M_{\rm limit}=\max[10^{13.4} h^{-1} M_{\odot}, M_f]$, with
$M_f$ corresponding to the mass related to the flux-limit at the
different redshifts, estimated by solving Eq.\,(\ref{bolom}) and
Eq.\,(\ref{sz}) for $M$. In the lower panels we present the
fractional difference between the two time-varying vacuum models
($\Lambda_{t}$CDM and $\Lambda_{t}G_{t}$CDM) and the $\Lambda$CDM,
similarly to Fig.\ref{figs:fig2}, but now for the realistic case of
the previously mentioned future cluster surveys. The error bars
shown correspond to 2$\sigma$ Poisson uncertainties, which however
do not include cosmic variance and possible observational systematic
uncertainties, that would further increase the relevant variance.

\begin{figure}[t]
\begin{center}
\mbox{\epsfxsize=14.5cm \epsffile{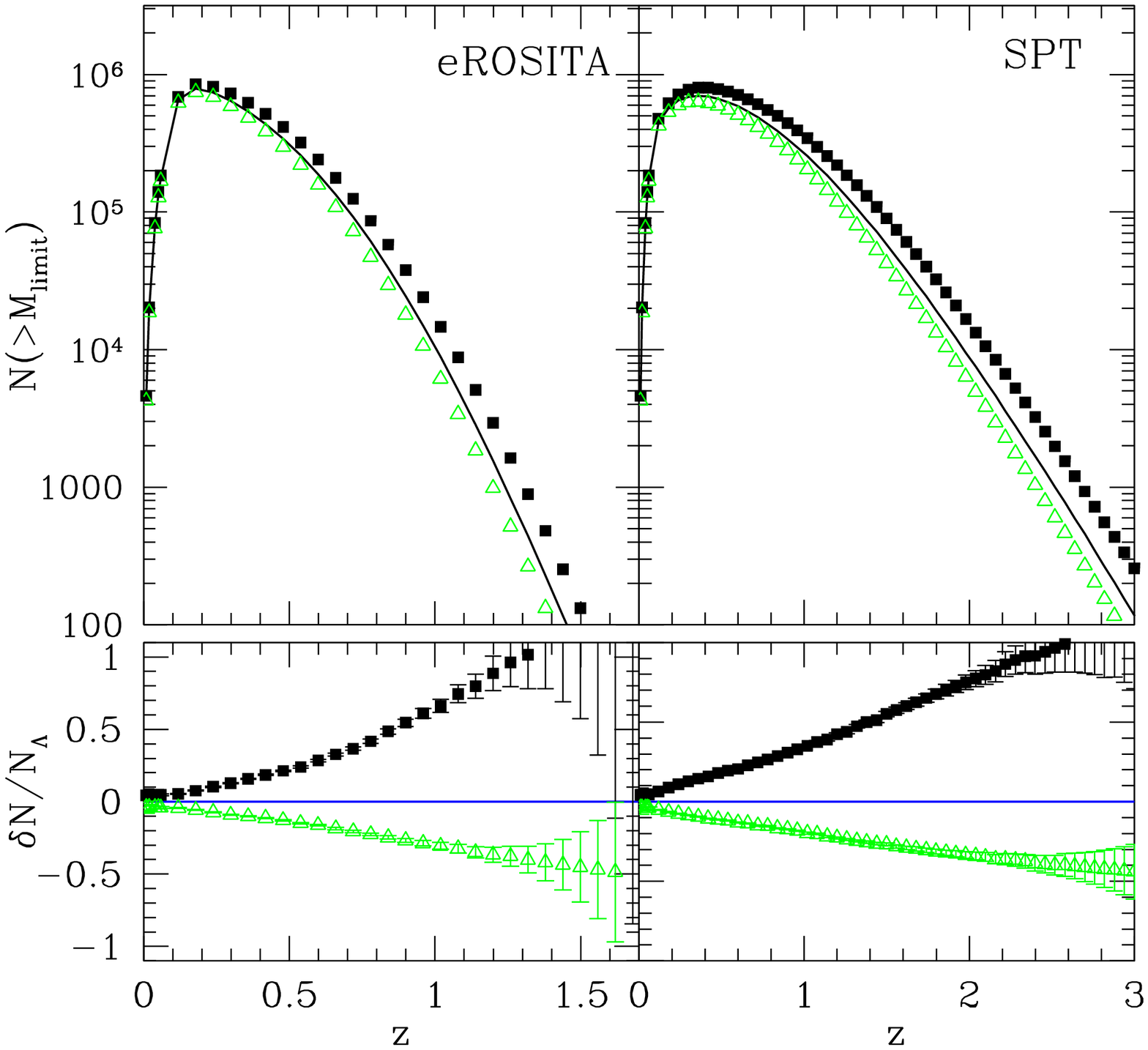}} \caption{Extension of
the analysis presented in Fig.\,4, but for the $\Lambda_{t}G_{t}$CDM
model only. Once more the theoretical redshift distribution of
$M\magcir 10^{13.4} h^{-1} M_{\odot}$ clusters in the two future
surveys is shown in the upper panel, whereas the corresponding
fractional difference with respect to the $\Lambda$CDM model is
shown in the lower panel. In this case, we use the limiting values
of $\nu$ comprised within the $1\sigma$ range of our fit, namely
$\nu=-0.004$ (black squares) and $\nu=+0.002$ (green triangles). See
Table 1.}\label{figs:fig3a}
\end{center}
\end{figure}

It is evident that the imposed flux limits, together with the
scarcity of high-mass halos at large redshifts, induces an abrupt
decline of ${\cal N}(z)$ with $z$, especially in the case of the
{\tt eROSITA} X-ray survey (note the shallower redshifts depicted in
Fig.\ref{figs:fig3} with respect to Fig.\ref{figs:fig2}).
%Furthermore, the uncertainties indicate
%that significant differences should be expected to be measured up to $z\simeq 1$ for the
%case of the {\tt eROSITA} X-ray survey, and up to $z\simeq 2$ for the
%case of the South Pole Telescope SZ survey, with the largest and most significant
%relative deviations occurring at $z\sim 0.8$ and $\sim 1.6$,
%respectively for the two surveys.
Furthermore, in the case of the $\Lambda_{t}$CDM model the
uncertainties indicate that we should expect significant and large
differences to be measured in both experiments for any redshift
$\magcir$ 0.3-0.4. On the other hand, the deviations of the
$\Lambda_{t}G_{t}$CDM model with respect to the concordance model,
although significant, are very small (for the central fit values)
and thus more difficult to detect.

%In  the lower panels of Fig. 3 we display the relative differences of
%the vacuum models but only up to a redshift at which they are
%significant, that is, such that:
%\be\label{stat} \frac{{\cal
%N}_{\rm VC}-{\cal N}_{\Lambda}}{ ({\cal N}_{\rm VC}-{\cal
%N}_{\Lambda})^{1/2} }> 3.5,
%\ee
%with ${\cal N}_{\rm VC}$ the redshift
%distribution predicted by either the
%$\Lambda_{t}$CDM or $\Lambda_{t}G_{t}$CDM
%model and ${\cal
%N}_{\Lambda}$ the corresponding redshift distribution of
%$\Lambda$CDM model. However, such a criterion does not take into
%account cosmic variance and possible observational systematic
%uncertainties which can hamper detecting small (but according to
%Eq. \ref{stat} significant) relative differences. We believe that
%relative differences of $\mincir 5\%$ will be difficult to detect
%especially at relative high redshifts.
 \begin{table*}[t]
\tabcolsep 3pt \vspace {0.2cm}
\begin{tabular}{|lc|cccc|cc|cc|} \hline \hline
Model     & symbol & $\Omega_m^0$& $\nu$& $\sigma_{8}$
&$\delta_{c}$& \multicolumn{2}{|c}{($\delta{\cal N}/{\cal
N}_{\Lambda})_{\rm eROSITA}$} &
\multicolumn{2}{|c|}{$(\delta {\cal N}/{\cal N}_{\Lambda})_{\rm SPT}$} \\
  &  & & & & & $z<0.2$ & $0.5\le z <1$ &  $z<0.2$ & $1 \le z <2$ \\ \hline
$\Lambda$CDM      & {\scriptsize black line}     & 0.284 & 0 & 0.811 & 1.675 & 0.00& 0.00 & 0.00 & 0.00\\
$\Lambda_{t}G_{t}$CDM &{\scriptsize black crosses}& 0.283 & -0.001 &0.818 & 1.677& 0.01 &0.04&  0.01 & 0.05\\
$\Lambda_{t}$CDM &{\scriptsize red dashed line}   & 0.284 &
-0.0017&0.840 &1.685 & 0.05 &0.25 & 0.06 & 0.36\\ \hline

$\Lambda_{t}G_{t}$CDM &{\scriptsize black squares}   & 0.283 & -0.004 & 0.853 &1.683& 0.06 &0.32 &  0.07 & 0.45\\
$\Lambda_{t}G_{t}$CDM &{\scriptsize green triangles} & 0.283 &
0.002 &0.786  &1.671& -0.05 &-0.18 &  -0.05 & -0.23\\ \hline
\end{tabular}
\caption[]{Numerical results. The $1^{st}$ column indicates the dark
energy model. The $2^{nd}$ column shows the symbols or lines used to
represent the different models in Figs. \ref{figs:fig2},
\ref{figs:fig3} and \ref{figs:fig3a}. The $3^{rd}$, $4^{th}$ and
$5^{th}$ columns list the central values of the input fitted
cosmological parameters (first three rows, one for each model being
considered) and their limiting $1\sigma$ values (last two rows, only
for the $\Lambda_{t}G_{t}$CDM model) obtained by using
%for the cosmological parameters in our fit to
the SNIa+BAO+CMB data (cf. section
\ref{sec:ConfrontingRunningFLRW}). The procedure to compute the
critical overdensity $\delta_c$ for each model ($6^{th}$ column) is
explained in Appendix ~\ref{sec:app}. The remaining columns present
the overall, within the indicated redshift bins, fractional relative
difference
[$\delta\mathcal{N}/\mathcal{N}_\CC\equiv(\mathcal{N}-\mathcal{N}_\CC)/\mathcal{N}_\CC$]
between the cluster abundances predicted by the time varying vacuum
models and by the $\Lambda$CDM cosmology, for the two future cluster
surveys discussed in the text. No errors are shown since the
2$\sigma$ Poisson uncertainties are less than $10^{-2}$.}
\end{table*}

Let us stress that the previous analysis of the cluster-size halo
redshift distribution has focused on the central values of our fits
to the SNIa+BAO+CMB data (cf. section
\ref{sec:ConfrontingRunningFLRW}). However, there is no compelling
reason to prefer the central values to any other value within the
1$\sigma$ contour. This observation is particularly important when
the dispersion turns out to be considerably large in that range. As
we saw in section 6, the $1\sigma$ range of the fitted $\nu$
parameter for both running models is indeed quite large and includes
both positive and negative values. In particular, for the
$\Lambda_{t}G_{t}$CDM model we have obtained $\nu\in [-0.004,
0.002]$. This model deserves special attention in that, in contrast
to the $\Lambda_{t}$CDM, the time evolution of the vacuum energy
density is compatible with matter conservation (a feature that
shares with the standard $\CC$CDM model). It is therefore worthwhile
to perform a more exhaustive investigation of the cluster halo
redshift distribution for this model within the limits of the
$1\sigma$ range, so as to check if the predicted deviations with
respect to the $\Lambda$CDM can be significant. We display the
results of this extended analysis in Fig.\,\ref{figs:fig3a}, where
we show, following a pattern entirely similar to that in Fig.
\ref{figs:fig3}, the expectations of the $\Lambda_{t}G_{t}$CDM model
for the limiting $\nu$ values in its $1\sigma$ range [ie.,
$\nu=-0.004$ (black squares) and $\nu=0.002$ (open triangles)]. It
is evident that, in this case, the predicted redshift distribution
of clusters, ${\cal N}(z)$, for the $\Lambda_{t}G_{t}$CDM model does
show significantly larger, and potentially measurable, differences
($\sim 50\%$) with respect to the concordance $\Lambda$CDM model,
even at moderate redshifts.

In Table 1, we show a more compact presentation of our results,
including the relative fractional difference between the ``running''
vacuum models and the $\Lambda$CDM model, in two characteristic
redshift bins, and for both future surveys under consideration ({\tt
eROSITA} and SPT). The first redshift bin corresponds to the local
universe ($z\mincir 0.2$) and it is shown in order to allow us to
judge whether the existing cluster samples could be used to
discriminate among the different models.
%The second redshift bin is that where the largest and most
%significant model deviation occur, according to the previous discussion.
It is evident that in the local range $z\mincir 0.2$, although there
are deviations among the models, their amplitude is very small and
it will be rather difficult to detect them unambiguously. However,
at larger redshifts,
%especially around the
%optimal redshift of each model (i.e., $z\sim 0.8$ and $\sim 1.6$ for
%the X-ray and SZ cluster surveys, respectively),
the amplitude of the deviations is such that it should be relatively
straightforward to distinguish among the different models.
%choose among the different ``running'' vacuum models.

To conclude, in view of the results of our analysis of the cluster
halo redshift distribution, ${\cal N}(z)$, presented in Figures
\ref{figs:fig3}, \ref{figs:fig3a} and Table 1, we can say that both
running models show large and significant differences with respect
to the $\Lambda$CDM within the 1$\sigma$ range of the fitted values
of the fundamental parameter $\nu$ of these models. We have
demonstrated that these deviations could be measured for the two
future cluster surveys considered in our study ({\tt eROSITA} and
SPT) provided we focus on sufficiently high cosmological redshifts
$z\magcir 0.3$, therefore outside the local domain. In practice,
{\tt eROSITA} will be most efficient in the redshift range
$0.3\lesssim z\lesssim 1.2$ whereas SPT in the wider segment
$0.3\lesssim z\lesssim 2.2$, where in both cases the upper limit on
$z$ is determined by the sharp decrease in the cluster number counts
and the corresponding larger statistical error.

Let us finally point out that, had we used the same rms mass
fluctuation normalization for both time-varying vacuum models, i.e.,
$\sigma_{\rm 8}=\sigma_{8, \Lambda}$ (as in \cite{BPS09a}), we would
have found that none of the time-varying vacuum models could be
distinguished from the reference $\Lambda$CDM model at a significant
level. This implies that the value of $\sigma_8$ plays also a
fundamental role in these kind of studies, especially when the
models have similar Hubble functions and perturbation growing modes,
as it is the case in the presently studied time-varying vacuum
models. Therefore, our including the computation of the value of
$\sigma_8$ for each of the models under consideration has been an
important feature of our study, and illustrates the general need to
follow this practice for other cosmological models.

%Finally we would like to mention that an interesting paper
%appeared recently \cite{suman10} and
%among other issues, it compares different forms
%of the halo mass function and its redshift evolution using N-body
%simulations of the $\Lambda$CDM and $w$CDM
%($w=$const) models. They do find some differences at the few percent level.
%Although our analysis is
%self-consistent, in the sense that we compare the expectations of DE
%models with respect to those of the concordance cosmology using the
%same mass function model, we plan
%to investigate in a forthcoming paper how sensitive are our
%observational predictions
%to the different mass functions fitting formulas.

\section{Conclusions}
In this work, we have analyzed the observational status of the
``running FLRW model'' (denoted by $\Lambda_{t}G_{t}$CDM), i.e. the
cosmological model characterized by the vacuum energy density
evolving quadratically with the expansion rate,
$\rL(H)=n_0+n_2\,H^2$ (with $n_0\neq 0$). The non-vanishing
coefficient $n_2$ provides the time evolution of the cosmological
term, and since it is a dimensional quantity it is conveniently
parametrized as $n_2=(3\nu/8\pi)\,M_P^2$, where $\nu$ plays the
role of the dimensionless $\beta$-function of the running $\rL$.
This is particularly clear also from the simultaneous running of the
gravitational coupling in this model, which evolves logarithmically
with the expansion rate according to $G^{-1}(H)\sim M_P^2(1+\nu\ln
H^2)$. These two running laws are intimately correlated such that
the Bianchi identity is automatically satisfied, as demanded by
general covariance. As a result, a nice feature of this model is
that matter is locally and covariantly conserved, similarly to the
standard $\CC$CDM model. It means that, in the $\Lambda_{t}G_{t}$CDM
model, there is no decay of vacuum energy into matter or vice versa.
This is in contradistinction to a previous version of the model, the
running $\Lambda_{t}$CDM model, which was recently confronted with
the latest observations in \cite{BPS09a}. We have re-analysed this
model here, together with the $\Lambda_{t}G_{t}$CDM model, so as to
better highlight the similarities as well as the important
differences between them. In both running models, $\rL(H)$ is
varying formally in the same way as a function of $H$, but in the
$\Lambda_{t}$CDM model $G$ is strictly constant at the price of
permitting a continuous exchange of energy between vacuum and
matter.

We have confronted $\Lambda_{t}G_{t}$CDM and $\Lambda_{t}$CDM  in
the light of the latest high-quality cosmological data from distant
type Ia supernovae, baryonic acoustic oscillations and the cosmic
microwave background anisotropies. This has allowed us to put a
limit on the size of the fundamental parameter $\nu$ controlling the
running of the vacuum energy in both models. This parameter also
controls the running of the gravitational coupling in the
$\Lambda_{t}G_{t}$CDM case, or alternatively the exchange of vacuum
energy and matter in the $\Lambda_{t}$CDM model. Despite the fact
that the two models are qualitatively quite different, for both of
them we find that if $\nu<0$ the formation of structure is
reinforced with respect to the $\CC$CDM, whereas for $\nu>0$ it is
depleted. This is understandable in the sense that, for $\nu<0$, the
vacuum energy decreases in the past (it may even become negative)
and therefore the formation of structure is favored. In contrast,
for $\nu>0$ the vacuum energy becomes larger and positive in the
past, preventing the growth of structure.

Although the central fitted $\nu$ values of the two running models
are different, in both cases we meet the preferred sign $\nu<0$,
with a magnitude of order $|\nu|={\cal O}(10^{-3})$. Furthermore,
the corresponding $1\sigma$ ranges are quite consistent among them,
especially when one takes into account their relative uncertainty,
which is of the same order as the central value, i.e.
$|\delta\nu|={\cal O}(10^{-3})$. As a result, both the
$\Lambda_t$CDM and $\Lambda_{t}G_{t}$CDM models could show
significant departures with respect to the $\CC$CDM power spectrum
within one standard deviation of the fitted values. This fact
translates then into a measurable impact on the redshift
distribution of cluster-size halos, as we have verified in detail,
finding that both running models could lead to very important
deviations with respect to the concordance model.

Interestingly enough, the redshift distribution of cluster-size
halos will be measured by two future important surveys, one is based
on the X-ray {\tt eROSITA} satellite and the other on the
Sunyaev-Zeldovich observations performed by the South Pole Telescope
(SPT).
%For both running models we
%find that for the most negative allowed {\bf value (within the $1\sigma$
%uncertainty range)} of the fundamental
%parameter $\nu$,
Our analysis shows that by sampling $\nu$ within its $1\sigma$ range
there is a significant maximal deviation --positive or negative,
depending on the sign of $\nu$-- in the predicted abundance of
clusters with respect to the $\Lambda$CDM, which could amount to an
anomaly of $50\%$ or higher at redshifts ranges ($z\gtrsim 0.3$)
which should be perfectly accessible to both realistic surveys.

Finally, let us emphasize that both models display an effective
equation of state behavior which can mimic both quintessence and
phantom dark energy, without of course involving quintessence or
phantom scalar fields. It follows that the sole EoS behavior of the
running cosmologies can be highly distinctive with respect to the
concordance $\CC$CDM model and it can also be used to distinguish
between the two running models themselves. This strategy  should be
most efficient upon combining the effective EoS determination with
the predicted deviations of the clustering redshift distribution
with respect to the $\CC$CDM. The upshot of our study is that the
``running cosmologies'' could provide an alternative and successful
version of dynamical dark energy which should be testable in the
next generation of cosmological experiments.

\vspace{0.5cm}

\acknowledgments Authors JG and JS have been partially supported by
DIUE/CUR Generalitat de Catalunya under project 2009SGR502; JS also
by MEC and FEDER under project FPA2010-20807 and by the
Consolider-Ingenio 2010 program CPAN CSD2007-00042. Author SB wishes
to thank the Dept. ECM of the Univ. de Barcelona for the
hospitality, and the financial support from the Spanish Ministry of
Education, within the program of Estancias de Profesores e
Investigadores Extranjeros en Centros Espanoles (SAB2010-0118). MP
acknowledges funding by Mexican CONACyT grant 2005-49878. JS would
like to thank Julio C. Fabris for discussions and the Brazilian
agency CNPq, and the Univ. Federal do Espiritu Santo, Brazil, for
the financial support and the warm hospitality extended to him while
doing part of this work.

\appendix
\section{The critical overdensity in time-varying vacuum models}\label{sec:app}
In this appendix we explain in detail the computation of
$\delta_{c}$ (the linearly extrapolated density threshold above
which structures collapse) for the RG models we have
considered\footnote{We follow the standard methods available in the
literature -- see e.g. \cite{Pace10} and \cite{Abramo}, and
references therein. Here we just extend them to encompass the class
of time varying $\rL(t)$ and $G(t)$ models under consideration.}.
The quantity $\delta_{c}$ is used for the study of the halo
abundances and their evolution in
Sect.\,\ref{sec:RunningFLRWCollapse}.

\subsection*{The $\Lambda_{t}G_{t}$CDM model:}

First we have to derive the second-order differential equation
governing the non-linear evolution of the matter perturbations in
our model. Using the Newtonian formalism for the cosmological fluid,
we start by writing the continuity, Euler and Poisson equations in
the matter dominated epoch:
\begin{eqnarray}
  \frac{\partial\rho_m}{\partial t}+\nabla_{\vec{r}}\cdot(\rho_m\vec{v})= 0
\label{eqn:cnpert}\;,\\
  \frac{\partial\vec{v}}{\partial
    t}+(\vec{v}\cdot\nabla_{\vec{r}})\,\vec{v}+
  \nabla_{\vec{r}}\,\Phi=0\label{eqn:enpert}\;,\\
  \nabla^2\Phi=4\pi G_{\!N}\sum_i\rho_i(1+3\omega_i)\;,\label{eqn:pnpert}
\end{eqnarray}
where $\vec v$ is the total velocity of the co-moving observer in
three-space, $\Phi$ is the Newtonian gravitational potential,
$\vec{r}$ is the physical coordinate, $G_{\!N}$ is the Newton's
coupling and $\sum_i$ runs over all the energy components, in our
case non-relativistic matter and the running cosmological constant
($i=m,\,\Lambda$). Finally, $\omega_i=p_i/\rho_i$ is the EoS
parameter for each component ($0$ for dust, and $-1$ for the CC,
respectively). Let us recall that, within this framework, $G_{\!N}$
and $\rho_\Lambda$ depend on time (at the background level). We
introduce comoving coordinates $\vec{x}=\vec{r}/a$ and define the
perturbations in the following way:
\begin{eqnarray}
\rho_i(\vec{x},t) & = & \bar{\rho}_i(t)+\delta\rho_i(\vec{x},t)=\bar{\rho}_i(t)(1+\delta_i(\vec{x},t))\;, \label{eqn:rpert} \\
\Phi(\vec{x},t) & = & \Phi_0(\vec{x},t)+\phi(\vec{x},t)\;, \label{eqn:fpert}\\
 \vec{v}(\vec{x},t) & = & a(t)[H(t)\vec{x}+\vec{u}(\vec{x},t)]\;, \label{eqn:vpert}\\
G_N(\vec{x},t) & = & G(t)+\delta G(\vec{x},t)\;.\label{eqn:gpert}
\end{eqnarray}
Here $H(t)$ is the Hubble function, and $\vec{u}(\vec{x},t)$ is the
comoving peculiar velocity. We have introduced also a perturbation
for $G_N$, equation (\ref{eqn:gpert}), which is mandatory in order
to have a consistent picture in the model under consideration
\cite{GSFS10}. Furthermore, in view of the corresponding EoS for
vacuum and non-relativistic matter, $\delta
p_{\CC}=-\delta\rho_{\CC}$ and $\delta p_m=0$. Our next task is to
insert Eqs.~(\ref{eqn:rpert})--(\ref{eqn:gpert}) into
Eqs.~(\ref{eqn:cnpert})--(\ref{eqn:pnpert}). To this end we first
use the definition of the gradient with respect to co-moving
coordinates,
$\vec{\nabla}\equiv\nabla_{\vec{x}}=a(t)\nabla_{\vec{r}}$,
%{\setlength\arraycolsep{1.4pt}
%\begin{eqnarray}
%\nabla_{\vec{x}}&\equiv&\vec{\nabla}=a(t)\nabla_{\vec{r}}\\
%\left(\frac{\partial}{\partial
%t}\right)_{\vec{x}}&\equiv&\frac{\partial}{\partial t}=
%\left(\frac{\partial}{\partial
%t}\right)_{\vec{r}}+H\left(\vec{x}\cdot\vec{\nabla}\right)\,,
%\end{eqnarray}}
and in this way we can express the result as follows:
\begin{eqnarray}
 \dot{\delta}_m+(1+\delta_m)\vec{\nabla}\cdot\vec{u} & = & 0\;, \label{eq:pertCont}\\
 \frac{\partial \vec{u}}{\partial t}+2H\vec{u}+(\vec{u}\cdot\vec{\nabla})\vec{u}+\frac{1}{a^2}\vec{\nabla}\phi & = & 0\;,\label{eq:pertEuler} \\
 \nabla^2\phi-4\pi G a^2\left(\bar{\rho}_m\delta_m-2\bar{\rho}_\Lambda\delta_\Lambda\right)-4\pi a^2\delta G\left(\bar{\rho}_m-2\bar{\rho}_\Lambda\right) & = & 0\;. \label{eq:pertPois}
\end{eqnarray}
Note that in order to get (\ref{eq:pertCont}) we have assumed the
condition $\vec{\nabla}\delta_m=0$, which holds for the spherical
collapse of a top-hat distribution\,\cite{Abramo}. Next we take the
divergence of the Euler equation (\ref{eq:pertEuler}) while using
the following identity\,\footnote{Note that in
(\ref{eq:decomposition}) we have assumed vanishing shear and
rotation tensors\,\cite{Pace10} owing to the assumed spherical
symmetry with a top-hat profile.}
\begin{equation}
 \vec{\nabla}\cdot[(\vec{u}\cdot\vec{\nabla})\,\vec{u}] =
\frac{1}{3}(\vec{\nabla}\cdot\vec{u})^2\;, \label{eq:decomposition}
\end{equation}
together with the time derivative of the continuity equation
(\ref{eq:pertCont}). Combining all three equations, we finally
obtain the fully non-linear evolution of the matter density
contrast:
\begin{equation}
\ddot{\delta}_m+2H\dot{\delta}_m-\frac{4}{3}\frac{\dot{\delta}_m^2}{1+\delta_m}-4\pi
G(1+\delta_m)\left(\bar{\rho}_m\delta_m-2\bar{\rho}_\Lambda\delta_\Lambda\right)-4\pi\delta
G(1+\delta_m)\left(\bar{\rho}_m-2\bar{\rho}_\Lambda\right)= 0
\label{nonl1}\;.
\end{equation}
Inserting now Eq.~(\ref{correlation}) into (\ref{nonl1}), we are
left with:
\begin{equation}
\ddot{\delta}_m+2H\dot{\delta}_m-\frac{4}{3}\frac{\dot{\delta}_m^2}{1+\delta_m}-4\pi
G\bar{\rho}_m(1+\delta_m)\left[\delta_m+\frac{\delta G}{G}\right]=
0\;. \label{nonl11}
\end{equation}
Changing the independent variable from cosmic time $t$ to the scale
factor $a$ through the relation $\partial_t=aH\partial_a$, equation
(\ref{nonl11}) can be rewritten
\begin{equation}
\label{nonldef}
\delta_m^{\prime\prime}+\left(\frac{3}{a}+\frac{H^\prime}{H}\right)\delta_m^\prime-\frac{4}{3}\frac{\delta_m^{\prime
2}}{1+\delta_m}-\frac{3\tilde{\Omega}_m(a)}{2a^2}(1+\delta_m)\left[\delta_m+\frac{\delta
G}{G}\right]\;,
\end{equation}
where $\tilde{\Omega}_m(a)$ is defined in (\ref{Omegastilde}) and
$f'=\partial_a f$ for any $f$. In the particular case when $G$ is
constant there are no perturbations of $G$ in (\ref{eqn:gpert}) and
then equation (\ref{nonldef}) boils down to Eq.~(18) in
\cite{Pace10}, as it should (see also Eq.~(7) in \cite{Abramo}).
Indeed, for constant $G$, the relation (\ref{Omegastilde}) tells us
that
$\tilde{\Omega}_m(a)=\Omega_m(a)/E^2(a)=\OMo/\left(a^3\,E^2(a)\right)$.
Let us also remark that (\ref{nonldef}) reduces to
\begin{equation}\label{lin}
 \delta_m^{\prime\prime}+\left(\frac{3}{a}+\frac{H^\prime}{H}\right)\delta_m^\prime-
 \frac{3\tilde{\Omega}_{\mathrm{m}}(a)}{2a^2}\left(\delta_m+\frac{\delta G}{G}\right)=0\;,
\end{equation}
if we keep only the linear terms. The latter is formally identical
to equation (\ref{difff1}) of Sect.\,\ref{sec:runningFLRW}.

Next, following the prescriptions of \cite{Pace10}, we compute
$\delta_c(a_f)$ (for any scale factor $a_f$, in particular for
$a_f=a_0=1$) by numerically integrating Eqs.~(\ref{nonldef}) and
(\ref{lin}) in the following manner:
\begin{enumerate}
\item First, we run the second order non-linear differential equation (\ref{nonldef})
between $a_i$ and $a_f$ (where $a_i$ is a sufficiently small scale
factor, which we take as $10^{-6}$). Our aim is to find the initial
value $\delta_m(a_i)$ for which the collapse takes place at $a=a_f$,
i.e. such that $\delta_m(a_f)$ is very large (formally infinite) at
the collapsing time. In practice (in order to set the initial
conditions), we may assume that the collapse is achieved once
$\delta_m$ is sufficiently large and set e.g. $\delta_m(a_f)=10^7$.
This coincides with the value chosen by \cite{Pace10}, although we
have checked that our results remain practically the same if we take
different large values for $\delta_m(a_f)$, say $10^5$ or $10^9$. As
for the initial condition on $\delta_m'$, and since we know that at
$a=a_i$ this derivative should very small, we can take (once more
following \cite{Pace10}) $\delta_m'(a_i)=5\cdot 10^{-5}$. Again, any
other small number (including 0) would yield virtually the same
results.

\item Second, we use the value for $\delta_m(a_i)$ computed in the
first step (together with $\delta_m'(a_i)=5\cdot 10^{-5}$) as the
initial condition for the linear equation (\ref{lin}). Solving the
latter for $\delta_m(a_f)$ we find $\delta_c(a_f)$ by definition of
this quantity (the linearly extrapolated density threshold above
which structures collapse).
\end{enumerate}

Before proceeding with the above two-step procedure, we have to get
rid of $\delta G$ on the  \textit{r.h.s.} of (\ref{nonldef}) and
(\ref{lin}) as follows. Since the growth factor in a pure matter
(Einstein-de Sitter, i.e. CDM) universe evolves as $D_{\rm EdS}=a$,
we normalize our growth factor such as to get $D\simeq a$ at early
enough epochs due to the dominance of the non-relativistic matter
component. Thus we first solve the third-order linear equation
(\ref{supeq}) between $a_i$ and $a_f$. We take $D(a_i)=a_i$,
$D'(a_i)=1$ and $D''(a_i)=0$ as the initial conditions, and then use
the solution to construct an interpolation function for $\delta
G(a)$ through (\ref{correlation}) [in order to do that we assume
$\delta G(a_i)=0$]. Using this function $\delta G(a)$ as an input
for Eqs.(\ref{nonldef}) and (\ref{lin}) we finally proceed as
explained in steps 1) and 2) above. The typical behavior of $\delta
G(a)$ can be seen in \cite{GSFS10}, Fig.~3b.

In the left panel of Fig.~\ref{fig6} we present the evolution of
$\delta_{c}(z)$ in the $\Lambda_t G_t$CDM model, using the different
values of $\nu$ that have been considered throughout this paper. The
corresponding values at the present epoch (effectively defining the
collapse time) are indicated by $\delta_{c}\equiv \delta_c(0)$. They
were used in the number counts analysis of
Sect.\ref{sec:HaloMasFunction}, and are listed in Table 1. In the
left panel of Fig.~\ref{fig6} we have also included $\delta_c(z)$
for the standard model $\CC$CDM (for which $\delta_c(0)=1.675$), and
also the constant value for the CDM model\,\footnote{For spherical
collapse, the CDM result is known and can be computed exactly:
$\delta_c=\frac{3}{20}(12\,\pi)^{2/3}\simeq 1.686$.}.

\FIGURE[t]{\centering \mbox{\epsfxsize=7.4cm \epsffile{fig6a.eps}}
\mbox{\epsfxsize=7.4cm \epsffile{fig6b.eps}} \caption{The critical
overdensity $\delta_c(z)$ as a function of the redshift for the
different models considered in the paper. The CDM and $\Lambda$CDM
determinations of $\delta_c(z)$ (with $\Omega_m^0=0.284$) are
represented by the dashed gray and solid black lines respectively.
For the $\Lambda_t G_t$CDM model, we study the case where we allow
for perturbations in $\rho_\Lambda$ and $G$ (left panel), with
$\Omega_m^0=0.283$ and three different values of $\nu$
($\nu=-0.004$, black squares; $\nu=-0.001$, black crosses;
$\nu=0.002$, green triangles) -- cf. Table 1 of
Sect.\,\ref{sec:HaloMasFunction}. Furthermore, for this model we
also consider the situation where these perturbations are neglected
(right panel). Finally, the dashed red line in the left panel
corresponds to the best fit value for the $\Lambda_t$CDM model (i.e.
$\Omega_m^0=0.284$, $\nu=-0.0017$). The precise values $\delta_c(0)$
for the two RG models with different inputs are also collected in
Table 1 of Sect.\,\ref{sec:HaloMasFunction}}\label{fig6}}

For comparison, we have also computed $\delta_{c}(z)$ by neglecting
the perturbations in $G(a)$ and $\rho_\Lambda(a)$ ($\delta
G=\delta_\Lambda=0$). This would correspond to the canonical or
simplest approach (i) mentioned in subsection
\ref{sec:RunningFLRWCollapse1}. Although the variable $G(a)$ and
$\rho_\Lambda(a)$ affect now the dynamics only at the background
level, they do influence non-trivially the evolution of the matter
perturbations through the modified Hubble function. Indeed, in this
approach, we have to solve equations (\ref{nonldef}) and (\ref{lin})
for $\delta G=0$, with $H(a)$ given by (\ref{Eg}) -- where both $G$
and $\rL$ are variable. Alternatively, we may use formally the same
approach as in\,\cite{Pace10} (in which $\delta G=0$ and  $G=G_0$)
if we employ our equations (\ref{DEpicture})-(\ref{rhoD}) and the
non-trivial effective EoS given by (\ref{weffCGCDM}). The
corresponding solution for $\delta_{c}(z)$ is displayed in the right
panel of Fig.~6. We can see that there are some differences with
respect to the results obtained in the left panel for non-vanishing
perturbations of $\rL$ and $G$, but the numerical deviations are not
dramatic (at the few per mil level).

\subsection*{The $\Lambda_{t}$CDM model:}
As before, we want to derive the second-order differential equation
governing the non-linear evolution of the matter perturbations. In
this case, however, the continuity equation involves exchange of
energy between matter and vacuum:
\begin{eqnarray}
  \frac{\partial\rho_m}{\partial t}+\nabla_{\vec{r}}\cdot(\rho_m\vec{v})=
-\dot{\rho}_\Lambda \label{eqn:cnpert2}\;.
\end{eqnarray}
The Euler and Poisson equations (\ref{eqn:enpert}) and
(\ref{eqn:pnpert}) remain unchanged, although we should remark that
the Newton's coupling $G_N=G_0$ is now strictly constant. As before,
we introduce comoving coordinates and define the perturbations as in
(\ref{eqn:rpert}) -- (\ref{eqn:vpert}), but now (consistently with
the approach we took in the paper) we will neglect the potential
perturbations in the running cosmological constant density
$\rho_\Lambda$ [so in (\ref{eqn:rpert}) we set $i=m$]. Inserting
Eqs.~(\ref{eqn:rpert})--(\ref{eqn:vpert}) into
Eqs.~(\ref{eqn:cnpert2}), (\ref{eqn:enpert}) and (\ref{eqn:pnpert})
we have:
\begin{eqnarray}
 \dot{\delta}_m+(1+\delta_m)\vec{\nabla}\cdot\vec{u} & = &-Q(t)\delta_m\;, \label{eq:pertCont2}\\
 \frac{\partial \vec{u}}{\partial t}+2H\vec{u}+(\vec{u}\cdot\vec{\nabla})\vec{u}+\frac{1}{a^2}\vec{\nabla}\phi & = & 0\;,\label{eq:pertEuler2} \\
 \nabla^2\phi-4\pi G_0 a^2\bar{\rho}_m\delta_m & = & 0\;, \label{eq:pertPois2}
\end{eqnarray}
where $Q(t)$ was defined in Sect.\,\ref{sec:runningLCDM}, see
equation (\ref{massvac}). By taking the divergence of the Euler
equation (\ref{eq:pertEuler2}) [using once more the identity
(\ref{eq:decomposition})], the time derivative of the continuity
equation (\ref{eq:pertCont2}) and combining all three equations, we
arrive at the fully non-linear evolution equation:
\begin{equation}
\label{eq:11t}
\ddot{\delta}_m+\left(2H+Q\right)\dot{\delta}_m-\frac{4\dot{\delta}_m^2+5Q\delta_m\dot{\delta}_m+Q^2\delta_m^2}{3(1+\delta_m)}+\left[2HQ+\dot{Q}-4\pi
G_0\bar{\rho}_m(1+\delta_m)\right]\delta_m=0\,,
\end{equation}
where we also used the background continuity equation
(\ref{Bronstein}). As expected, in the linear regime (\ref{eq:11t})
reduces to equation (\ref{eq:11}) of Sect.\,\ref{sec:runningLCDM}:
\begin{equation}
\label{eq:112}
\ddot{\delta}_{m}+(2H+Q)\dot{\delta}_{m}-\left[4\pi\,G_0{\bar{\rho}_{m}}-2HQ-\dot{Q}
\right]\delta_{m}=0\,.
\end{equation}
It is convenient to change the independent variable from $t$ to $a$
and perform  the numerical integration of (\ref{eq:11t}) and
(\ref{eq:112}) following the two-step procedure 1) and 2) described
in detail above.
%\begin{equation}
%\label{eq:11}
%\delta_m''+\left(\frac{3}{a}+\frac{H'}{H}+\frac{Q}{aH}\right)
%\delta_m'-\frac{1}{3(1+\delta_m)}\left(4\delta_m^{\prime 2}+5\frac{Q}
%{aH}\delta_m\delta'_m+\frac{Q^2}{(aH)^2}\delta_m^2\right)-\left[\frac{3}{2}\tilde{\Omega}_{\mathrm{m}}(a)(1+\delta_m)-\frac{2Q}{H}-\frac{a}{H}Q'\right]\frac{\delta_m}{a^2}=0\,.
%\end{equation}
In particular, equation (\ref{eq:112}) can be rewritten in the scale
factor variable as
\begin{eqnarray} \label{eq:112a}
\delta_m''+\left(\frac{3}{a}+\frac{H'}{H}+\frac{Q}{aH}\right)\delta_m'
&-&\left[\frac{3}{2}\tilde{\Omega}_{m}(a)-\frac{2Q}{H}-\frac{a}{H}Q'\right]\frac{\delta_m}{a^2}=0\,,
\end{eqnarray}
where again $f'=\partial_a f$ for any quantity $f$ in this equation.
For the particular case of time independent vacuum energy
($\dot{\rho}_{\Lambda}=0$) we have $Q=Q'=0$, and then upon using
$\tilde{\Omega}_m(a)=\OMo/\left(a^3\,E^2(a)\right)$ we see that
(\ref{eq:112a}) boils down to equation (19) of Ref.\,\cite{Pace10},
as it should. The solution of (\ref{eq:112a}), as a part of the
aforementioned two-step procedure, provides $\delta_c$ as a function
of the scale factor or as a function of the redshift $z=(1-a)/a$, as
shown in the left panel of Fig.\,\ref{fig6} (see the dashed red line
in that figure). For the numerical analysis we used the inputs
$\Omega_{m}^{0}=0.284$ and $\nu=-0.0017$, as indicated in Table 1.
The corresponding result for $\delta_c\equiv\delta_c(z=0)$ can be
directly read off Fig.\,\ref{fig6}, and is quoted in that table.

From the range of $\delta_c$ values obtained in Table 1 for the two
models, we see that they vary from $1.671$ to $1.685$, and hence
correspond to variations of $-2.4$ per mil and $+5.9$ per mil,
respectively, compared to the $\CC$CDM model value
($\delta_c(0)=1.675$).

%%%%%%%%%%%%%%%%%%%%%%%%%%%%%%%%%%%%%%%%%%%%%%%%%%%%%%%%%%%%%%%%%%%%%%%%%
%\newcommand{\JHEP}[3]{{ J. of High Energy Physics } {JHEP} {#1} (#2)  {#3}}
\newcommand{\JHEP}[3]{ {JHEP} {#1} (#2)  {#3}}
\newcommand{\NPB}[3]{{ Nucl. Phys. } {\bf B#1} (#2)  {#3}}
\newcommand{\NPPS}[3]{{ Nucl. Phys. Proc. Supp. } {\bf #1} (#2)  {#3}}
\newcommand{\PRD}[3]{{ Phys. Rev. } {\bf D#1} (#2)   {#3}}
\newcommand{\PLB}[3]{{ Phys. Lett. } {\bf B#1} (#2)  {#3}}
\newcommand{\EPJ}[3]{{ Eur. Phys. J } {\bf C#1} (#2)  {#3}}
\newcommand{\PR}[3]{{ Phys. Rep. } {\bf #1} (#2)  {#3}}
\newcommand{\RMP}[3]{{ Rev. Mod. Phys. } {\bf #1} (#2)  {#3}}
\newcommand{\IJMP}[3]{{ Int. J. of Mod. Phys. } {\bf #1} (#2)  {#3}}
\newcommand{\PRL}[3]{{ Phys. Rev. Lett. } {\bf #1} (#2) {#3}}
\newcommand{\ZFP}[3]{{ Zeitsch. f. Physik } {\bf C#1} (#2)  {#3}}
\newcommand{\MPLA}[3]{{ Mod. Phys. Lett. } {\bf A#1} (#2) {#3}}
%%%%%%%%%%%%%%%%%%%%%%%%%%%%%%%%%%%%%%%%%%%%%%%%%%%%%%%%%%%%%%%%%%%%%%%%%
%%%%%%%%%%%%%%%%%%%%%%%%%%%%%%%%%%%%%%%%%%%%%%%%%%%%%%%%%%%%%%%%%%%%%%%%%
\newcommand{\CQG}[3]{{ Class. Quant. Grav. } {\bf #1} (#2) {#3}}
\newcommand{\JCAP}[3]{{ JCAP} {\bf#1} (#2)  {#3}}
\newcommand{\APJ}[3]{{ Astrophys. J. } {\bf #1} (#2)  {#3}}
\newcommand{\AMJ}[3]{{ Astronom. J. } {\bf #1} (#2)  {#3}}
\newcommand{\APP}[3]{{ Astropart. Phys. } {\bf #1} (#2)  {#3}}
\newcommand{\AAP}[3]{{ Astron. Astrophys. } {\bf #1} (#2)  {#3}}
\newcommand{\MNRAS}[3]{{ Mon. Not. Roy. Astron. Soc.} {\bf #1} (#2)  {#3}}
\newcommand{\JPA}[3]{{ J. Phys. A: Math. Theor.} {\bf #1} (#2)  {#3}}
\newcommand{\ProgS}[3]{{ Prog. Theor. Phys. Supp.} {\bf #1} (#2)  {#3}}
\newcommand{\APJS}[3]{{ Astrophys. J. Supl.} {\bf #1} (#2)  {#3}}
%%%%%%%%%%%%%%%%%%%%%%%%%%%%%%%%%%%%%%%%%%%%%%%%%%%%%%%%%%%%%%%%%%%%%%%%%

\newcommand{\Prog}[3]{{ Prog. Theor. Phys.} {\bf #1}  (#2) {#3}}
\newcommand{\IJMPA}[3]{{ Int. J. of Mod. Phys. A} {\bf #1}  {(#2)} {#3}}
\newcommand{\IJMPD}[3]{{ Int. J. of Mod. Phys. D} {\bf #1}  {(#2)} {#3}}
\newcommand{\GRG}[3]{{ Gen. Rel. Grav.} {\bf #1}  {(#2)} {#3}}

%  Example:  \NPB {\bf 20} {1992}  {200}

%%%%%%%%%%%%%%%%%%%%%%%%%%%%%%%%%%%%%%%%%%%%%%%%%%%%%%%%%%%%%%%%%%%%%%%%%
%\newpage
%%%%%%%%%%%%%%%%%%%%%%%%%%%%%%%%%%%%%%%%%%%%%%%%%%%%%%

\end{document}